\documentstyle[preprint,eqsecnum,aps]{revtex}
\tightenlines

\newcommand{\C}{\mbox{$\,${\sf I}\hspace{-1.2ex}{\bf C}}}
\newcommand{\D}{\mbox{$\,$/\hspace{-1.4ex}D}}
\begin{document}
\def\be{\begin{equation}}
\def\ee{\end{equation}}
\def\mn{\mu\nu}
\def\ha{{1\over 2}}
\def\astf{{}^*f_{\mn}}
\def\ore{\overrightarrow{e}}
\def\orv{\overrightarrow{v}}
\def\orh{\overrightarrow{h}}
\def\ora{\overrightarrow{a}}
\def\appD{\buildrel\approx\over{D}}
\def\rappD{\overrightarrow\appD}
\def\lappD{\overleftarrow\appD}
\def\simD{\buildrel\sim\over{D}}
\def\simG{\buildrel\sim\over{G}}
\def\Ft{\buildrel\sim\over{F}}
\def\p{\phi}
\def\ph{\hat\phi}
\def\Uem{U(1)_{e.m.}}
\def\gp{\bar g_\phi}
\def\Dt{\buildrel\sim\over{\D}}
\def\Gt{\buildrel\sim\over{\Gamma}}
\def\Dtt{\buildrel\approx\over{\D}}
\def\rsimD{\overrightarrow{\simD}}
\def\lsimD{\overleftarrow{\simD}}
\def\B{{\cal B}}
\def\F{{\cal F}}
\def\P{{\cal P}}
\def\L{{\cal L}}
\def\H{{\cal H}}
\def\W{{\cal W}}
\def\M{{\cal M}}
\def\R{{\cal R}}
\def\pa{\partial}
\def\tn{\tilde\nabla}
\def\a{\alpha}
\def\d{\delta}
\def\l{\lambda}
\def\r{\rho}
\def\s{\sigma}
\def\ga{\gamma}
\def\Ga{\Gamma}
\def\om{\omega}
\def\ka{\kappa}
\def\si{\sigma}
\def\b{\beta}
\def\w{\mbox{w}}
\def\vphi{\varphi}
\def\fnm{\phi^{(n,m)}}
\def\te{\tilde\eta}
\def\ha{\frac{1}{2}}
\def\vv{\vphi}
\def\ve{\varepsilon}
\def\lp{l_\vphi}
\def\px{\psi(x)}
\def\P{{\cal P}}
\def\H{{\cal H}}
\def\hpss{{\hps}_{s_3}}
\def\Us{U^{(s)}}
\def\Vpm{V^\pm_m}
\def\Mpm{{\cal M}^\pm_m}
\def\ua{\underline{a}}
\def\Gam{\Gamma_\mu}
\def\da{\dagger}
\def\gat{\tilde\gamma}
\def\pr{\psi_R}
\def\pl{\psi_L}
\def\dtt{\buildrel\approx\over\delta}
\def\gt{\tilde g}
\def\gtt{\tilde g'}
\def\gz{\tilde g_0}
\def\e{\tilde e}
\def\eh{{e\over{\hbar c}}}
\def\LW{\buildrel\approx\over{\cal L}_{W_4}}
\def\LWo{\buildrel\approx\over{\cal L}}
\def\q{\hat q}
\def\tw{\theta_W}
\def\hph{\hat\vphi_0}
\def\WW{\hat W}
\def\ZZ{\hat Z}
\draft 
\preprint{MPI-PhT/98-68}
\title{Mass-Generation by Weyl-Symmetry Breaking}
\author{Wolfgang Drechsler\footnote{wdd@mppmu.mpg.de}}
\address{Max-Planck-Institut f\"ur Physik\\
         F\"ohringer Ring 6, 80805 M\"unchen, Germany}
\maketitle
\vskip-1.0cm
\begin{abstract}
\renewcommand{\baselinestretch}{1.0}
\small\normalsize
A massless electroweak theory for leptons is formulated in a Weyl space,
$W_4$, yielding a Weyl invariant dynamics of a scalar field $\phi$, chiral
Dirac fermion fields $\psi_L$ and $\psi_R$,
and the gauge fields $\kappa_\mu, A_\mu,
Z_\mu, W_\mu$ and $W_\mu^\dagger$ allowing for conformal rescalings of the
metric $g_{\mu\nu}$ and all fields with nonvanishing Weyl weight together
with the corresponding transformations of the Weyl vector fields, $\kappa_\mu$,
representing the $D(1)$ or dilatation gauge fields. The local group
structure of this Weyl-electroweak (WEW) theory is given by
$G=SO(3,1)\otimes D(1)\otimes\simG$ --- or its universal covering group
$\bar G$ for the fermions --- with $\simG$ denoting the
electroweak gauge group $SU(2)_W\times U(1)_Y$. In order to investigate the
appearance of nonzero masses in the theory the Weyl-symmetry is explicitly
broken by a term in the Lagrangean constructed with the curvature scalar
$R$ of the $W_4$ and a mass term for the scalar field.
Thereby also the $Z_\mu$- and $W_\mu$- gauge fields as well as the charged
fermion field (electron) acquire a mass as in the standard electroweak theory.
The symmetry breaking is governed by the relation $D_\mu\Phi^2=0$, where
$\Phi$ is the modulus of the scalar field and $D_\mu$ denotes the
Weyl-covariant derivative. This true symmetry reduction, establishing a
scale of length in the theory by breaking the $D(1)$ gauge symmetry,
is compared to the so-called spontanous symmetry breaking in the
standard electroweak theory which is, actually, the choice of a
particular (nonlinear) gauge obtained by adopting an origin, $\hat\phi$,
in the coset space representing $\phi$ with $\hat\phi$ being invariant under
the electromagnetic gauge group $U(1)_{e.m.}$. Particular attention is devoted
to the appearance of Einstein's equations for the metric after the
Weyl-symmetry breaking yielding a pseudo-Riemannian space, $V_4$, from
a $W_4$ and a scalar field with a constant modulus $\hat\vphi_0$. The quantity
$\hat\vphi_0^2$ affects Einstein's gravitational constant in a manner comparable
to the Brans-Dicke theory. The consequences of the broken WEW theory
are worked out and the determination of the parameters of the theory
is discussed.
\end{abstract}
\newpage
\section{Introduction}
The standard model in particle physics was very successful in giving
a unified description of strong, weak and electromagnetic processes.
However, it was always considered a problem by many theorists
to understand properly the {\it mass giving algorithm} in this
theory, i.e.\ to answer the question what lies behind the
so-called ``Higgs phenomenon''. Is this so-called spontanous
symmetry breaking by the vacuum
expectation value of a scalar quantum field --- yielding at the same
time the gauge boson and the fermion masses in the theory --- the
correct way to account for the appearance of nonzero masses
in physics? In other words, is the Higgs mechanism of the
standard model only a convenient algorithm to generate masses
of the gauge and fermion fields without loosing the renormalizability
of the theory, or has this enigmatic phenomenon a deeper understanding 
with the present formulation of the theory yielding only a 
{\it particular parametrization} of a so far not very well understood
aspect of the theoretical description: the origin of nonzero masses
in nature? Related to this is the question whether the scalar
quantum field $\p$ needed to formulate the Higgs mechanism is, indeed, a true
spin zero matter field which materializes as a spinless particle
of a certain mass showing up in high energy processes.

Recently there have appeared several suggestions of a different
nature and interpretation for the mass-giving mechanism in 
particle physics which would not require a Higgs particle 
to exist. The proposal of Pawlowski and R\c{a}czka (see \cite{1} and
references therein) stresses the conformal invariance of the
original theory without actually breaking the conformal symmetry
in the course of introducing masses, but rather considering the
appearance of nonzero mass ratios as a certain choice of gauge.
The other method which was proposed by the present author in
collaboration with H. Tann \cite{2} aims at accounting
for the origin of nonzero masses by using a Weyl-geometric
framework starting from an originally massless Weyl-symmetric
theory and a subsequent explicit breaking of the Weyl symmetry with the help
of a term in the Lagrangean involving the curvature scalar $R$ of the
Weyl space $W_4$ and the mass of the scalar field. In this framework
the mass generation in an originally massless and scaleless theory is
considered to be due to an interplay between the ambient
geometry --- a Weyl-geometry --- and a universal scalar quantum field.
The breaking $W_4\to V_4$ yielding a pseudo-Riemannian, i.e.\
$V_4$, description in the limit is formulated as a condition on the
Weyl vector fields $\ka_\r$ which in turn
represent an aspect of the Weyl-geometry (see Sect. II
below) being given in the broken case
as a derivative of the scalar field
yielding thus, finally, zero length curvature, i.e.\ $f_{\mn}=0$.
[The Weyl vector fields $\ka_\r$ are the $D(1)$ or
dilatation gauge potentials and $f_{\mu\nu}$ are
the corresponding curvature components.] At the same time the
validity of Einstein's equations for the metric in the $V_4$ limit is
required to be satisfied with the energy-momentum tensors of the
now massive fields as sources and with --- as it turns out ---
a gravitational coupling
constant depending on the modulus $\Phi$ of $\p$.
On the one hand, i.e.\ as far as gravitation is concerned,
the squared modulus, $\Phi^2$, of the scalar field plays the
role of a Brans-Dicke-type field in this
formalism \cite{3}. On the other hand, i.e. as far as the generation
of nonzero masses for the gauge and fermion fields is concerned,
the field $\p$ plays a role of a Higgs-type field in this
broken Weyl theory.

The idea of our geometrically motivated method basically
is that in accounting for the
origin of nonzero masses in nature the theoretical framework
should include gravitation from the outset starting thus from the
investigation of a dynamics of massless boson and fermion fields
formulated in a Weyl space $W_4$ containing the dynamics of
a metric modulo conformal rescalings. Subsequently the Weyl-symmetry
is broken explicitly yielding a Riemannian description together
with a definite length (and mass) scale being established and
a set of field equations for the metric being required to be satisfied.

The plan of the paper is as follows. After some introductory remarks
on Weyl geometry and Weyl spaces we briefly review the theory
presented in \cite{2} in which the geometry of a Weyl space of
dimension four and its use in elementary particle theory was studied 
in detail. In this work the scalar field --- called $\vv$ there ---
was a complex quantum field with nonzero Weyl weight carrying
besides its transformation character under Weyl transformations
[see below] no further representation properties. This Weyl
covariant theory is then generalized in Sect. II~by including
the electroweak gauge group $\simG = SU(2)_W\times U(1)_Y$ in
the description with the scalar field --- now denoted by $\p$ ---
possessing in addition representation properties with respect
to the weak isospin group $SU(2)_W$ as well as the hypercharge
group $U(1)_Y$. After some general remarks about symmetry breaking
in gauge theories at the end of Sect. II, this Weyl-electroweak
theory (WEW theory) is explicitly broken in Sect. III~and the
role of electromagnetism (Subsection A) and gravitation
(Subsection B) in the resulting theory, formulated in a Riemannian
space, is investigated in detail. Subsection C deals with
the wave equation for the scalar field $\p$, and Subsection D, finally, is
devoted to the determination of the free parameters of the theory.
An essential point in the
presented discussion is that the breaking of the Weyl-symmetry
and the appearance of a length and mass scale in the theory
is qualitatively different from the so-called spontanous
symmetry breaking in the electroweak theory yielding the masses
of the gauge boson and fermion fields there. We show, using the
coset representation of the scalar field $\p$ derived and
discussed in Appendix A and B, that the so-called spontanous
symmetry breaking is a particular choice of gauge within
the electroweak theory which is characterized by the choice of
an origin $\ph$ in the coset space, with $\ph$ being invariant
under the {\it electromagnetic} gauge group $\Uem$ exhibiting
thus, finally, a residual $\Uem$ gauge symmetry of the theory
which is, in fact, a nonlinear realization of the original
$\simG$-symmetry on the subgroup $\Uem$. We end in Sect. IV~with 
some concluding remarks and discussions of the results obtained
due to the true symmetry breaking occurring in this
Weyl-electroweak theory when a definite intrinsic unit of lengths
is established.


\section{Electroweak Theory in Weyl-Symmetric Form}

\noindent{\bf A. Geometric Preliminaries.}

In this section we investigate a unified electroweak
theory in the presence of ``gravitation'' formulated in a Weyl space.
Since Einstein's metric theory of gravitation is neither
conformally invariant nor a theory which could be formulated in
a Weyl space we have to break the Weyl-symmetry in a second step,
as mentioned, in order to recover Einstein's theory in this framework.
On the other hand, we want to generate masses for the matter fields of
the theory by starting from a gauge dynamics involving at first only massless
fields in order to see how this mass generation obtained by
Weyl-symmetry breaking compares to the Higgs mechanism in the
standard model. To facilitate our discussion we shall, however,
disregard the $SU(3)$ gauge group of colour and shall treat
only the electroweak part of the standard model together with gravitation
neglecting thus the strong interactions. Moreover, we treat only
one generation of fermions for simplicity, i.e. the leptons
$e$ and $\nu =\nu_e$. Our main interest here is to see how the
electron mass $m_e$ and the $SU(2)$-gauge boson masses
appear due to Weyl-symmetry breaking.
The fact that a second fermion generation with $m_\mu > m_e$ introduces
{\it another} and {\it different} mass scale given by the myon mass
$m_\mu$ --- opening up, moreover, the possibility of $\mu_{e3}$-decay ---
and, similarly, for the third generation with a tau-mass
$m_\tau > m_\mu$, cannot be explained by the present model.
In this respect the broken Weyl theory is, unfortunately,
not better than the standard model.

In order to stress the role of gravitation in the present context,
we like to make the following remarks. The usual Higgs
mechanism yielding nonzero masses for the fields in the standard model
does not limit the actual size of the masses obtained: The mass value
for the fermion fields could be shifted to arbitrary large values.
Of course, the gauge boson masses are related to the strength of
the Fermi coupling constant for charge changing weak currents.
However, the feature that the elementary fermion masses could be
arbitrary large seems to be an unphysical one since the generation of nonzero
rest masses is accompanied by the generation of gravitational fields
and gravitational interactions. If gravity were included in the
standard model one would expect in a consistent theory that elementary
masses cannot be shifted to arbitrary large values due to
damping effects resulting from the consideration of
gravitational interactions. This is reminiscent of Hermann Weyl's
remark that a theory which tries to account for the
origin of masses in nature cannot be formulated consistently
without considering gravitation at the same time. The standard
model should thus be formulated in a general relativistic setting.
In starting from a massless conformally invariant scenario
a formulation of the original dynamics in a Weyl space
or Weyl geometry would thus be very suggestive. Let us, therefore,
begin our investigation by formulating a gauge dynamics of a massless
spin zero and a single generation of massless spin $\ha$
fields in a Weyl space $W_4$.

In \cite{2} the geometry of a Weyl space was investigated
in detail. We shall use the notation defined there (see in particular
Appendix A of this paper). We shall refer to \cite{2} as to I in
the following and refer, for example, to Eq.(1.2) of I
as to (I, 1.2) etc..

A Weyl space $W_4$ is characterized by two differential forms:
\be
ds^2= g_{\mu\nu}(x) dx^\mu\otimes dx^\nu\/;\qquad
\/\ka =\ka_\mu (x)dx^\mu \,.
\label{21}
\ee
A $W_4$ is equivalent to a family of Riemannian spaces
\be
(g_{\mn}, \ka_\si),\/(g'_{\mn}, \ka'_\r),\/(g''_{\mn}, \ka''_\r)\ldots
\label{22}
\ee
with metrics $g_{\mn}(x),~\/ g'_{\mn}(x)\ldots$ and Weyl vector fields
$\ka_\r (x),\ka'_\r (x)\ldots$ related by
\begin{eqnarray}
g'_{\mu\nu}(x)&=&\si(x)\, g_{\mu\nu} (x)  \\ \label{23a}
\kappa'_\r(x)&=&\kappa_\r(x) +\pa_\r \log\si (x)\,,\label{23b}
\end{eqnarray}
where $\si (x)\!\in\! D(1)$, $\si (x)=e^{\r(x)}>0$, with
$D(1)$ denoting the dilatation group which is isomorphic
to $R^+$ (the positive real line). The transformations 
(2.3) and (2.4) are called {\sl Weyl-transformations}
involving a conformal rescaling (2.3) of the metric
together with the transformation (2.4) of the Weyl
vector fields. In the following discussion we shall consider
a Weyl space of dimension $d=4$ possessing Lorentzian
signature $(+,-,-,-)$ of its metrics. (For reference to the
earlier history of Weyl spaces and Weyl geometry see the
references quoted in I.)

A $W_4$ reduces to a Riemannian space $V_4$ for
$\ka_\mu =0$; a $W_4$ is equivalent to a $V_4$ if the
``length curvature'' associated with the Weyl vector field
$\ka_\mu$ is zero, i.e.\ for
\be
f_{\mn}=\pa_\mu\ka_\nu -\pa_\nu\ka_\mu =0\\.
\label{24}
\ee

In I we studied a Weyl invariant dynamics of massless
fields involving the metric $g_{\mn}$, the Weyl vector or
$D(1)$ gauge fields $\ka_\r$, and the electromagnetic, i.e.\
$U(1)=U(1)_{e.m.}$\ gauge fields $A_\mu$, as well as the ``matter''
fields $\vphi$ (spin zero) and $\psi$ (spin $\ha$, Dirac spinor) with
Weyl weight $w(\vphi)=-\ha$ and $w(\psi)=-{3\over 4}$~[see I]. It
turned out in the discussion given in I that $\vphi$ is not
a bona fide matter field but could better be characterized
as a universal Bans-Dicke-type field or a Higgs-type field related
to symmetry breaking. On the other hand, the spinor field
$\psi$ {\it is} a true matter field representing
leptons in the present formalism.

The local group structure of the theory studied in I was
$SO(3,1)\otimes D(1) \otimes U(1)$ for the nonfermionic fields and
$Spin(3,1)\otimes D(1)\otimes U(1)$ for the Dirac
spinor field $\psi$ with $Spin(3,1)$ denoting the universal
covering group of the orthochronous Lorentz group 
$SO(3,1)\equiv O(3,1)^{++}$ acting in the local spin space $\C_4$
representing the standard fiber of the spinor bundle on
which the field $\psi(x)$ is defined as a section (see I
and the discussion below), and with $U(1)$ denoting the
electromagnetic gauge group.

The pull back of a connection on the corresponding frame bundle
was denoted by the one-forms of Weyl weight zero (Latin indices
are local Lorentzian indices):
\be
(w_{ik}=-w_{ki}\,,\ka\,, A)
\label{25}
\ee
with coefficients with respect to a natural base
$dx^\mu$ in the dual tangent space $T_x^*(W_4)$ to $W_4$ at $x$
given by:
\be
(\Gamma_{\mu ik}(x)=-\Gamma_{\mu ki}(x)\,,\ka_\mu (x)\,, A_\mu(x))\,.
\label{26}
\ee
The first entry in (\ref{25}) is Lorentz-valued (i.e.\ antisymmetric
in $i$ and $k$), the second is $D(1)$-valued (corresponding to a
real, noncompact, abelian gauge group), and the third is $U(1)$-valued 
(corresponding to the complex, compact, abelian electromagnetic
gauge group). The fully covariant derivative of a 
tensor quantity $\phi^{(n,m)}$ of type $(n,m)$, i.e.\ covariant
of degree $n$ and contravariant of degree $m$, possessing Weyl
weight $w(\phi^{(n,m)})$ and charge $q$, i.e.\ transforming under
Weyl transformations (2.3) and (2.4) as
\be
{\fnm}'(x)=[\si(x)]^{w(\fnm)}~\fnm (x)\,,
\label{27}
\ee
and under electromagnetic gauge transformations as
\be
{\fnm}'(x)=e^{-{iq\over\hbar c}}~\fnm (x)\,,
\label{28}
\ee
is given by
\begin{eqnarray}
\tilde D\fnm &=&D\fnm + {iq\over \hbar c} A\fnm\nonumber\\
              &=&\nabla\fnm -\om(\fnm)\kappa\fnm +{iq\over \hbar c}
                 A\fnm .\label{29}
\end{eqnarray}
Here $D = dx^\mu D_\mu$ denotes the Weyl-covariant derivative 
(I, A3), and $\ka =\ka_\mu dx^\mu$, $A=A_\mu dx^\mu$. Furthermore,
$\nabla =dx^\mu\nabla_\mu$ with $\nabla_\mu$ denoting the covariant
derivative with
respect to the Weyl-connection $\Gamma_{\mn}{}^\r$ defined by
\be
\Gamma_{\mn}{}^\r=\bar\Gamma_{\mn}{}^\r +W_{\mn}{}^\r =\ha g^{\r\l}
    (\pa_\mu g_{\nu\l}+\pa_\nu g_{\mu\l}-\pa_\l g_{\mn})-
    \ha(\ka_\mu\d_\nu^\r +\ka_\nu\d_\mu^\r-\ka^\r g_{\mn})\/.
\label{210}
\ee
Here and in the sequel purely metric quantities pertaining to a
$V_4$ are denoted by a bar, for example, 
$\bar\Ga_{\mn}{}^\r=\{{\r\atop\mn}\}$ are the Christoffel
symbols of the metric $g_{\mn}$. We remark in
passing that the connection coefficient $\Ga_{\mn}{}^\r$,
defined with respect to a natural base in (\ref{210}), is
Weyl-invariant, obeying $\Ga_{\mn}{}^\r =\Ga_{\mn}{}^\r {}\,'$, with
the change in the metric computed according to (2.3) being compensated
by the change in the Weyl vector fields computed according to (2.4).
Thus the Weyl-covariant derivative $D\phi^{(n,m)}$ of a quantity
$\phi^{(n,m)}$ is independent of the Weyl gauge chosen in the
family (2.2), and, by definition, transforms again
like (\ref{27}). Corresponding to Eq.(\ref{29}) the Weyl- and $U(1)$-
covariant derivative of a spinor field with $w(\psi)=-{3\over 4}$ and
charge $e$ is given by
\begin{eqnarray}
\tilde D\psi (x)&=&D\psi (x)+{ie\over\hbar c} A\cdot {\bf 1}\psi (x)\nonumber\\
&=&dx^\mu\biggl\{
\biggl(\pa_\mu+i\Gam (x)\biggr)\psi (x)+{3\over 4}\ka\cdot {\bf 1}~
\psi (x)+{ie\over\hbar c} A_\mu\cdot {\bf 1}~\psi (x)\biggr\}\,,\label{211}
\end{eqnarray}
where $\Gam (x)$ is the spin connection
\be
\Gam (x)=\l^j_\mu (x)\ha\Gamma_{jik}(x) S^{ik}~;\qquad S^{ik}=
{i\over 4}[\gamma^i,\gamma^k]\/.
\label{212}
\ee
Here $\ga^i;~i=0,1,2,3$ are the constant Dirac matrices satisfying
$\{\ga^i,\ga^k\}=\ga^i\ga^k+\ga^k\ga^i=2\eta^{ik}\cdot {\bf 1}$ with
$\eta^{ik}=diag(1,-1,-1,-1)$ and $\lambda_\mu^j (x)$ are
the vierbein fields obeying
\be
g_{\mn} (x)=\l^i_\mu (x)\l^k_\nu (x)\eta_{ik}\/.\label{213}
\ee
The inverse vierbein fields used below are denoted by $\l^\mu_i(x)$.
The quantities $\Ga_{\mu ik}=\l^j_\mu\Ga_{jik}$ appearing in
the first equation of (\ref{212}) are the coefficients of the
Lorentz part of the connection mentioned in relation to (\ref{25})
and (\ref{26}) with $\Ga_{jik}$ defined by
\be
\om_{ik}=\bar\om_{ik}-\ha(\ka_i\theta_k-\ka_k\theta_i)=\theta^j
\Ga_{jik}
\label{214}
\ee
where $\bar\om_{ik}=\theta^j\bar\Ga_{jik}$ with
$\bar\Ga_{jik}=-\bar\Ga_{jki}$ being the
Ricci rotation coefficients of a $V_4$ and 
$-\ha (\ka_i\eta_{kj}-\ka_k\eta_{ij})\theta^j$ denoting the
Weyl addition in a $W_4$. In Eq. (\ref{214}) $\theta^j=\l^j_\mu (x)dx^\mu$;
$j=0,1,2,3$ are the fundamental one-forms representing a
Lorentzian basis in the dual tangent space, $T_x^*(W_4)$, at
$x\in W_4$. (Compare Appendix A of I.)
The form (\ref{214}) for $\om_{ik}$ together with (\ref{210})
yields $D\l^\mu_i=0$ with $w(\l^\mu_i)=-\ha$
for the Weyl-covariant derivative of the vierbein field and,
correspondingly, $D_\r g_{\mn}=0$ with the Weyl weights 
$w(g_{\mn})=1$ [see Eq. (2.3)], $w(\l^i_\mu)=\ha$ and
$w(\eta_{ik})=0$ [compare (\ref{213})]. The relation $D_\r g_{\mn}=0$,
expressing the fact that the metric is Weyl-covariant constant,
reduces the connection on the general linear frame bundle [i.e.\ the
$Gl(4,R)$-bundle] in a Weyl space --- possessing a metric given
only modulo conformal transformations (2.3) --- to a Weyl frame
bundle, called $P_W$ in I, possessing the structural group
$SO(3,1)\otimes D(1)$.\\

\noindent{\bf B. Standard Model Extension: Weyl-Electroweak Theory (WEW Theory)}

We now extend the formalism developed in I to a unified electroweak
theory [neglecting as mentioned $SU(3)$ colour degrees of freedom]
by extending the gauge group of the theory to the group \cite{3a}
\be
G = SO(3,1)\otimes D(1)\otimes U(1)_Y\times SU(2)_W\/,
\label{215}
\ee
i.e.\ interpreting the $U(1)$ degree of freedom in I as
weak hypercharge, $U(1)_Y$, and considering an additional
weak isospin group $SU(2)_W$ (compare Weinberg's model of leptons \cite{4}).
The underlying principal bundle over
$W_4$ is now 
\be
P=P(W_4,G)\label{216}
\ee
with $G$ given by (\ref{215}). For the discussion of spinor fields
of Dirac type one considers, as usual, the spin frame bundle
$\bar P=\bar P (W_4,\bar G)$
possessing the universal covering group of $G$, i.e.
\be
\bar G = Spin (3,1)\otimes D(1)\otimes U(1)_Y\times SU(2)_W
\label{217}
\ee
as structural group (compare the discussion above and in I).
A Dirac spinor field $\psi$, with Weyl weight
$w(\psi)=-{3\over 4}$ as before [see I], and hypercharge $Y$,
possessing a definite representation character regarding
$SU(2)_W$ i.e.\ [we follow the standard model assignment]
$I=\ha$ (isodoublet), $Y=-\ha$ for the left-handed
chiral fields $\psi_L(x)=\ha(1-\ga_5)\psi (x)$, and  $I=0$
(isosinglet), $Y=-1$, for the right-handed chiral fields
$\psi_R(x)=\ha (1+\ga_5)\psi (x)$, with
\be
\ga_5 =i\ga_0\ga_1\ga_2\ga_3\,;\qquad \ga_5{}^\dagger =\ga_5\,;\qquad (\ga_5)^2=1\/.
\label{218}
\ee
The chiral fields $\psi_L (x)$ and $\psi_R (x)$ will be regarded,
respectively, as a section on the spinor bundle $S$ associated
to $\bar P$ with fiber $F$ given by $\tilde{\C}=\C_4\times \C_2$
for $I=\ha$; or given by $\tilde {\C}=\C_4\times {\bf \C}$ for
$I=0$, being thus defined by the bundle 
\be
S=S (W_4, F=\tilde{\C},\bar G)\,.
\label{219}
\ee
Hence the leptonic chiral fermion fields of Weyl weight $-{3\over 4}$
will be
\be
\psi_L={\nu_L\choose e_L}=
  {\ha (1-\ga_5)\psi_\nu\choose \ha(1-\ga_5)\psi_e}\,, Y=-\ha\,;
\quad \psi_R=e_R =\ha (1+\ga_5)\psi_e, Y=-1,
\label{220}
\ee
with their adjoints $(\bar\psi =\psi^\dagger\ga_0)$:
\be
\bar\psi_L=(\bar\nu_L,\bar e_L)=\left(\bar\psi_\nu\ha
(1+\ga_5),\bar\psi_e\ha (1+\ga_5)\right), 
Y=\ha;~\bar\psi_R=\bar e_R=\bar\psi_e\ha (1-\ga_5); Y=1\/.
\label{221}
\ee

For the scalar field we shall use as representation character
with respect to $SU(2)_W$ an isodoublet, $I=\ha$, yielding thus
\be
\phi = \left({\vphi_+\atop\vphi_0}\right) \quad {\rm with} \quad Y=\ha; 
\quad \mbox{and}\quad
\phi^\dagger =(\vphi^*_+,\vphi^*_0)\quad \mbox{with} \quad Y=-\ha
\label{222}
\ee
possessing the Weyl weight $w(\phi)=w(\phi^\dagger )=-\ha$. Here $\vphi_0$
is a neutral complex field, and $\vphi_+$
is a complex field with positive charge, obeying $\vphi^*_+=\vphi_-$.
The relation between electric charge $Q$, isospin $(I_3)$, and weak
hypercharge is, as usual,
\be
Q=I_3+Y\/.
\label{223}
\ee

The field $\phi$ may be regarded as a section on the bundle
\be
E=E(W_4, F=\C_2, G)\label{224}
\ee
associated to $P$. The square of the modulus of the scalar field is
now given by the $U(1)_Y$ and $SU(2)_W$ invariant of Weyl weight
$w(\Phi^2)=-1$:
\be
\Phi^2=\phi^\dagger \phi =\vphi^*_+\vphi_++\vphi^*_0\vphi_0=|\vphi_+|^2+
|\vphi_0|^2.
\label{225}
\ee

The invariant Yukawa coupling term of Weyl weight $-1$ for
the scalar and spinor fields --- reading
$\tilde\ga\sqrt{\vphi^*\vphi} (\bar\psi\psi)=
\tilde\ga\Phi(\bar\psi\psi)$ in I --- will now be written as
\be
\tilde\ga\{(\bar\psi_L\phi)\psi_R + \bar\psi_R (\phi^\dagger\psi_L)\}\/.
\label{226}
\ee

Calling the $U(1)_Y$ gauge potentials $B_\mu$, i.e.\ reserving as usual
the notation $A_\mu$ for the electromagnetic gauge potentials,
the full $G$-covariant derivative of $\phi$ is written as
[compare (\ref{215})]
\be
\buildrel\approx \over D_\mu\!\phi =D_\mu\phi +{i\over 2}\tilde g
A^a_\mu\tau_a\phi + i\tilde g'Y B_\mu\cdot {\bf 1}~\phi \/.
\label{227}
\ee
Here the Weyl-covariant part is given  by $D_\mu\phi=\pa_\mu\phi +\ha
\ka_\mu\cdot {\bf 1}\phi$, and the $SU(2)_W$-gauge fields are denoted by
$A^a_\mu;~a=1,2,3;\mu=0,1,2,3$. Moreover, $Y=\ha$ in
(\ref{227}) according to (\ref{222}). The Lie algebra
of $SU(2)_W$ adapted to the choice (\ref{222}) for
$\phi$ is given by $\ha\tau_a$ with $\tau_a; a=1,2,3$ denoting the
Pauli matrices [summation over $a$ from 1 to 3 is understood in
(\ref{227})]. Finally, $\tilde g$ and $\tilde g'$ are dimensionless
coupling constants for the $SU(2)_W$ and $U(1)_Y$ coupling, respectively,
which we write with a tilde in order not to confuse them with
the determinant of the metric tensor called $g$.

A similar expression as (\ref{227}) may be written down for
$\appD_\mu\!\psi_L$ involving the Weyl-covariant
part $D_\mu\psi_L =(\pa_\mu +i\Gamma_\mu)
\psi_L+{3\over 4}\ka_\mu\cdot {\bf 1} \psi_L$, an $SU(2)_W$ part as in (\ref{227}),
and an $U(1)_Y$ part with $Y=-\ha$. [Compare (\ref{211}).]
For $\appD_\mu\!\psi_R$ the $A^a_\mu$-contributions
are absent due to the choice $I=0$ for the right-handed fermion field.
We take account of this in the notation by writing only one tilde
for the covariant differentiation of $\psi_R$, i.e.
$\appD_\mu\!\psi_R\equiv\simD_\mu\!\psi_R=
D_\mu\psi_R+i\tilde g'Y B_\mu\cdot {\bf 1}\psi_R$ with $Y=-1$ according to
(\ref{220}).

We are now in a position to write down a $G$-gauge invariant
Lagrangean density of Weyl weight zero generalizing $\tilde\L_{W_4}$
of $I$ [compare (I, 3.8)] to the case of an
electroweak theory including ``gravitation'', i.e.\
containing also a dynamics for the metric (determined modulo
Weyl-transformations), in a scenario for the massless fields
$\phi,\psi_L,\psi_R,\ka_\mu, B_\mu$ and $A^a_\mu$ possessing
all a definit Weyl weight which is zero for the gauge fields
$\ka_\mu, B_\mu, A^a_\mu$, and is $w(\p)=-\ha$ and $w(\pl)=
w(\pr)=-{3\over 4}$ as mentioned above. Again we use below the
same coefficient for the kinetic term of the scalar field
and for the ${1\over{12}}R$-term implying in I the validity of
the relation (I, 2.20), while (I, 2.21) is a consequence
of the choice $w(\psi)=-{3\over 4}$. The Lagrangean for a massless
$G$-invariant theory, called for short WEW theory (Weyl-electroweak theory),
now reads:
\begin{eqnarray}	
\buildrel\approx\over{\cal L}_{W_4}&=&
K\sqrt{-g}\Biggl\{\ha g^{\mn}(\rappD_\mu\phi)^\dagger
\rappD_\nu\phi -{1\over {12}}
R\,\phi^\da\phi -\beta (\phi^\da\phi)^2+\tilde\a R^2
\nonumber\\ 
& &\qquad +{i\over 2}
\left(\bar\psi_L\ga^\mu{\rappD}_\mu\psi_L
-\bar\psi_L {\lappD}_\mu
\ga^\mu\psi_L\right)+{i\over 2}\left(\bar\psi_R\ga^\mu
\rsimD_\mu\psi_R-\bar\psi_R\lsimD_\mu\ga^\mu\psi_R\right)
\nonumber\\
& &
+\tilde\ga[(\bar\psi_L\phi)\psi_R+
\bar\psi_R(\phi^\dagger\psi_L)]-\tilde\delta
{1\over 4}f_{\mn}f^{\mn} -
\buildrel\approx\over\delta {1\over 4}\left(F_{\mn}^a
F^{\mn}_a + B_{\mn}B^{\mn}\right)\Biggr\}\/,\label{228}
\end{eqnarray}
Here $\ga^\mu$ denote a set of $x$-dependent $\ga$-matrices with
Weyl weight $w(\ga^\mu)=-\ha$ defined by
\be
\ga^\mu=\ga^\mu(x) =\l^\mu_i (x)\ga^i\quad\mbox{obeying}\quad
\{\ga^\mu,\ga^\nu\}=2 g^{\mn}\cdot {\bf 1}\/.
\label{229}
\ee

The meaning of the various terms in (\ref{228}) is the same as in I and was
described there in detail. This is true except for the generalization
of the covariant derivatives, as explained above, being denoted here
by $\appD_\mu$ and $\simD_\mu$ with the arrows $\rightarrow$ and $\leftarrow$
indicating, as usual, the action on $\pl$, $\pr$ and on $\bar\pl$, $\bar\pr$,
respectively, with a sign change involved according to the rule
[compare (\ref{211}) and (\ref{227})],
$\rappD_\mu{}^\dagger=\ga^0\lappD_\mu\ga^0$ for the fermion fields and
similarly for $\phi$, i.e.\ $({\rappD}_\mu\phi)^\dagger
=\phi^\dagger\lappD_\mu$ with
$\lappD_\mu ={\rappD}_\mu{}^\dagger = \appD_\mu{}\!^\dagger$. Corresponding
to this, i.e.\ to the introduction of the new gauge fields $B_\mu$ and
$A^a_\mu$, the last two terms in (\ref{228}) replace the 
electromagnetic term $-{1\over K}{1\over 4}F_{\mn} F^{\mn}$
in (I, 3.8) with $F_{\mn}=\pa_\mu A_\nu-\pa_\nu A_\mu$.
Since the fields $B_\mu$ and $A^a_\mu$ have the dimension
$[L^{-1}]$ ($L$=Length) as seen from (\ref{227}), we replace the
factor ${1\over K}$ (see below) appearing in front of the
electromagnetic term in (I, 3.8) by a constant $\buildrel\approx\over\d$
of dimension $[L^2]$. Moreover, the Lagrangean $\LW$ is chiral
invariant, i.e. is invariant under {\it global} $U(1)$ transformations
($\b ' = const$):
\be
\pl \rightarrow e^{-i\b '}~\pl \/;\quad \pr \rightarrow e^{i\b '}~\pr \/;
\quad \phi \rightarrow e^{-2i\b '}~\phi \/,
\label{230a}
\ee
and analogously for $\bar\pl$, $\bar\pr$ and $\phi^\da$ with the
complex conjugate phase factors.

The field strengths (curvatures) entering the expression (\ref{228})
in addition to $f_{\mn}$ and the Weyl curvature scalar $R$ defined by
[see (I, A31)]
\be
R=\bar R-3\bar\nabla^\r \ka_\r +{3\over 2} \ka^\r\ka_\r\,,
\label{230}
\ee
where $\bar R$ is the Riemannian part, are the $U(1)_Y$ gauge curvature
\be
B_{\mn}=\pa_\mu B_\nu -\pa_\nu B_\mu~,
\label{231}
\ee
and the $SU(2)_W$ i.e.\ Yang-Mills gauge curvature
\be
F^a_{\mn} =\pa_\mu A^a_\nu -\pa_\nu A^a_\mu-\tilde g f^a_{bc} A^b_\mu A^c_\nu\/,
\label{232}
\ee
with $f^a_{bc}=\ve_{abc}$ denoting the structure constants of
$SU(2)_W$ where $\ve_{abc}$ is the Levi-Civita symbol. The
overall constant $K$ in (\ref{228}) with dimension [Energy $\cdot L^{-1}$]
is a factor converting the length dimension in the curly brackets
(which is $[L^{-2}]$) into [Energy $\cdot L^{-3}$] in order to give
$\buildrel\approx\over{\L}_{W_4}$ --- finally, after symmetry breaking ---
the correct dimension of an energy density. This factor $K$ drops
out of the field equations in the Weyl-symmetric case discussed
in this section and appears in (\ref{228}) only for convenience.
We finally remark that the length dimension of the scalar field
$\p$ is assumed to be $[L^0]$ and relative to this choice the
fermion fields have length dimension $[L^{-\ha}]$. With this
convention the Yukawa coupling constant $\tilde\ga$ has length
dimension $[L^{-1}]$.

So far we have not included a coupling $\sim\sqrt{-g} F_{\mn} f^{\mn}$
in the Lagrangean (I, 3.8) or a coupling $\sim\sqrt{-g}B_{\mn}
f^{\mn}$ in (\ref{228}) which would also be of Weyl weight zero
and hence would be allowed to occur. We intend to come back to
an investigation of this point in a separate context and restrict the
discussion here to the direct product structure of the abelian
gauge groups involved, treating them thus as completely independent
from each another.

The variation of the fields in the Lagrangean (\ref{228}) now yields
the following set of $G$-covariant field equations:
\begin{eqnarray}
&&\delta\phi^\da :~~g^{\mn}\appD_\mu\appD_\nu \phi + {1\over 6}R\phi
 + 4\beta(\phi^\da\phi)\phi - 2\gat\bar\pr\pl = 0\/,\label{234}\\
&&\delta\pl^\da :~~-i\ga^\mu\appD_\mu\!\pl - \gat\phi\pr = 0\/,\label{235}\\
&&\delta\pr^\da :~~-i\ga^\mu\simD_\mu\!\pr - \gat(\phi^\da\pl) = 0\/,\label{236}\\
&&\delta\ka_\r :~~\tilde\d D_\mu f^{\mu\r} = -6\tilde\alpha D^\r R~,
\label{237}\\
&&\delta B_\r :~~\dtt D_\mu B^{\mu\r} = \gtt\left[j^{(\phi )}{}^\r+j^{(\pl )}{}^\r
                 +j^{(\pr )}{}^\r\right]\/,\label{238}\\
&&\delta A^a_\r :~~\dtt ~\appD_\mu\! F_a^{\mu\r}\equiv ~\dtt\left[D_\mu F_a^{\mu\r}
                   -\gt f^b_{ac}A^c_\mu F_b^{\mu\r}\right]=
                 \gt\left[j^{(\phi)}_a{}^\r + j^{(\pl)}_a{}^\r\right]\/,
\label{239}\\
&&\delta g^{\mn} :~~{1\over 6}\Phi^2\left[R_{(\mn)}-\ha g_{\mn}R\right]-
                    4\tilde\a R\left[R_{(\mn)}-{1\over 4}g_{\mn}R\right]-
                    4\tilde\a\left\{D_{(\mu}D_{\nu )}R-g_{\mn}D^\r D_\r R\right\}=
  \nonumber\\
&&\qquad\qquad =\Theta^{(\phi)}_{\mn} + T^{(\pl )}_{\mn} + T^{(\pr )}_{\mn}
                    + T^{(f)}_{\mn} + T^{(B)}_{\mn} + T^{(F_a)}_{\mn}-g_{\mn}\gat
                    \left[(\bar\pl\phi )\pr + \bar\pr (\phi^\da\pl)\right]\/.
\label{240}
\end{eqnarray}
Here we have used the following hermitean and $G$-gauge covariant
expressions for the weak hypercharge and isospin source currents:
\be
j^{(\phi)}_\r = {i\over 4}\left(\phi^\da\rappD_\r\phi -
                \phi^\da\lappD_\r\phi\right)
\label{241}
\ee
for the $U(1)_Y$ $\phi$-current (with $Y=\ha$ for the $\phi$-field),
\be
j^{(\pl)}_\r = -\ha(\bar\pl \ga_\r\pl)~; \qquad j^{(\pr)}_\r = -(\bar\pr\ga_\r\pr)\/,
\label{242}
\ee
for the left- and right-handed $U(1)_Y$ fermion currents (with $Y=-\ha$
and $Y=-1$, respectively), and for the $SU(2)_W$ currents
\be
j^{(\phi)}_{a\r} = {i\over 2}\left(\phi^\da\ha\tau_a\rappD_\r\phi -
                 \phi^\da\lappD_\r\ha\tau_a\phi\right)~,
\label{243}
\ee
\be
j^{(\pl)}_{a\r} = \bar\pl\ga_\r\ha\tau_a\pl~; \qquad j^{(\pr)}_{a\r}\equiv 0~.
\label{244}
\ee
The $G$-gauge covariant expressions for the symmetric energy-momentum
tensors appearing in (\ref{240}) are:
\begin{eqnarray}
\Theta^{(\phi)}_{\mn} &=& \ha\left[(\appD_\mu\!\phi)^\da(\appD_\nu\!\phi) +
                            (\appD_\nu\!\phi)^\da(\appD_\mu\!\phi)\right]
+{1\over 6}\left\{D_{(\mu}D_{\nu)}\Phi^2 - g_{\mn} D^\r D_\r\Phi^2 \right\}
\nonumber\\
&\hfil&\qquad\qquad\qquad\qquad\qquad
-g_{\mn}\left[\ha g^{\r\l}(\appD_\r\!\phi)^\da(\appD_\l\!\phi)
                   -\beta (\phi^\da\phi)^2\right]\/,
\label{245}\\
T^{(\pl)}_{\mn} &=& {i\over 2}\left\{\bar\pl \ga_{(\mu}\rappD_{\nu)}\pl -
\bar\pl\lappD_{(\mu}\ga_{\nu)}\pl\right\} - g_{\mn} {i\over 2}\left\{
\bar\pl\ga^\r\rappD_\r\pl - \bar\pl\lappD_\r\ga^\r\pl\right\}\/,
\label{246}
\end{eqnarray}
and analogously for $T^{(\pr)}_{\mn}$, and
\begin{eqnarray}
T^{(f)}_{\mn} &=&~-\,\tilde\delta\left[f_{\mu\s}f_\nu{}^\s - {1\over 4}
                   g_{\mn} f^{\r\l}f_{\r\l}\right]\/,
\label{247}\\
T^{(B)}_{\mn} &=& -\dtt\left[B_{\mu\s}B_\nu{}^\s - {1\over 4}
                    g_{\mn} B^{\r\l}B_{\r\l}\right]\/,
\label{248}\\
T^{(F_a)}_{\mn} &=& -\dtt\left[F^a_{\mu\s}F_{a\nu}{}^\s - {1\over 4}
                    g_{\mn} F_a^{\r\l}F^a_{\r\l}\right]\/.
\label{249}
\end{eqnarray}

While the last three energy-momentum tensors have vanishing trace
the traces of $\Theta^{(\phi)}_{\mn}, T^{(\pl)}_{\mn}$ and $T^{(\pr)}_{\mn}$
for solutions of the field equations (\ref{234}) -- (\ref{236}) and their
adjoints read:
\begin{eqnarray}
\Theta^{(\phi)\mu}_\mu &=& -{1\over 6}R\phi^\da\phi + \tilde\ga
\left[(\bar\pl\phi)\pr + \bar\pr(\phi^\da\pl)\right]\/,
\label{250}\\
T^{(\pl)\mu}_\mu &=& T^{(\pr)\mu}_\mu = {3\over 2}\tilde\ga
\left[(\bar\pl\phi)\pr + \bar\pr(\phi^\da\pl)\right]\/.
\label{251}
\end{eqnarray}
Using these relations in computing the trace of Eq. (\ref{240}),
together with $D^\r D_\r R=0$ following
from (\ref{237}), one finds that the resulting
equation is identically satisfied.

It is also interesting to compute the Weyl-covariant divergences of the
energy-momentum tensors (\ref{245}) -- (\ref{249}) for solutions of the
field equations, i.e. using Eqs. (\ref{234}) -- (\ref{239}) as well as the
Bianchi identities for the $F^a_{\mn}$ reading, with $\{\r\mn\}$ denoting
the cyclic sum of the indices $\r\mn$,
\be
\appD_{\{\r}F_{\mn\}}^a = 0\/.
\label{252}
\ee
The result is
\begin{eqnarray}
D^\mu\Theta^{(\phi)}_{\mn} &=& {1\over 6}\left[R_{(\nu\r)}-\ha g_{\nu\r}R\right]
D^\r\Phi^2-{1\over 12}\Phi^2 D^\r f_{\r\nu}+
\tilde\ga \left[(\bar\pl\pr\rappD_\nu\phi) + (\phi^\da\lappD_\nu\bar\pr\pl)\right]\/,
\label{253}\\
D^\mu T^{(\pl)}_{\mn} &=& \tilde\ga\, Y_\nu(\phi,\pl,\pr)+
\gtt B_{\nu\r}j^{(\pl)\r}+\gt F^a_{\nu\r}j^{(\pl)\r}_a\/,
\label{254}\\
D^\mu T^{(\pr)}_{\mn} &=& \tilde\ga\, Y_\nu(\phi,\pl,\pr)+
\gtt B_{\nu\r}j^{(\pr)\r}\/,
\label{255}
\end{eqnarray}
with the contribution $Y_\nu (\phi ,\pl,\pr)$ of the Yukawa coupling
to Eqs. (\ref{254}) and (\ref{255}):
\begin{eqnarray}
&&Y_\nu (\phi ,\pl,\pr) = {1\over 4}\left[
\left\{\bar\pl(\phi^\da\rappD_\nu\pl) + (\bar\pl\lappD_\nu\phi)\pr\right\}-
\left\{(\bar\pl\phi)\rsimD_\nu\pr + \bar\pr\lsimD_\nu (\phi^\da\pl)\right\}\right.
\nonumber\\
&&\qquad\qquad\left.
+\left\{(\pl\pr\rappD_\nu\phi) + (\phi^\da\lappD_\nu\bar\pr\pl)\right\}-
\left\{(\bar\pl\ga_\nu\ga_\r\pr\rappD{}^\r\phi) +
      (\phi^\da\lappD{}^\r\bar\pr\ga_\r\ga_\nu\pl)\right\}\right]\/,
\label{256}
\end{eqnarray}
and
\begin{eqnarray}
D^\mu T^{(f)}_{\mn} &=& 6\tilde\a f_{\nu\r}D^\r R\/,
\label{257}\\
D^\mu T^{(B)}_{\mn} &=& -\gtt B_{\nu\r}\left[j^{(\phi)\r}+j^{(\pl)\r}+j^{(\pr)\r}\right]\/,
\label{258}\\
D^\mu T^{(F_a)}_{\mn} &=& -\gt F^a_{\nu\r}\left[j^{(\phi)\r}_a + j^{(\pl)\r}_a\right]\/.
\label{259}
\end{eqnarray}

The last two equations as well as the field equations (\ref{238})
and (\ref{239}) show that it is useful to introduce the following
total $U(1)_Y$ and total $SU(2)_W$ current densities of Weyl weight zero
composed of bosonic and fermionic contributions:
\begin{eqnarray}
U(1)_Y~~:~~ J_\r &=& -{{\pa\LW}\over {\pa B^\r}} = K\sqrt{-g}\,\gtt
                \left[j^{(\phi)}_\r + j^{(\pl)}_\r + j^{(\pr)}_\r\right]\/,
\label{260}\\
SU(2)_W~:~~ J_{a\r} &=& -{{\pa\LW}\over {\pa A^{a\r}}} = K\sqrt{-g}\,\gt
                 \left[j^{(\phi)}_{a\r} + j^{(\pl)}_{a\r}\right]\/.
\label{261}
\end{eqnarray}
These definitions allow the separation of the Lagrangean density $\LW$
into a Weyl-invariant ``free'' part $\LWo{}\!^{(0)}_{W_4}$
(obtained from Eq. (\ref{228})
for $B_\r\equiv 0$ and $A^a_\r \equiv 0$) and a Weyl-invariant interaction part
$\LWo_{int}$ expressed in terms of the currents (\ref{260}) and
(\ref{261}) and the corresponding gauge fields, i.e.
\be
\LWo_{int}~= - J^\r B_\r - J^\r_a A^a_\r \/,
\label{261a}
\ee
yielding thus the decomposition
\be
\LW ~=~ \LWo{}\!^{(0)}_{W_4} + \LWo_{int} + \LWo{}\!^{(B,F_a)} \/,
\label{261b}
\ee
where $\LWo{}\!^{(B,F_a)}$ is given by the last two terms in (\ref{228})
proportional to $\dtt$ representing the Lagrangean density of the free
$B_\r$ and $A^a_\r$ fields.

The energy-momentum balance for the interacting fields is represented
by the Weyl-covariant divergence of Eq. (\ref{240}). Using Eqs. (\ref{253})
--(\ref{259}) together with the contracted Bianchi identities (I, A40)
for a $W_4$, it is seen that this is identically satisfied for solutions
of the field equations for any value of $\tilde\ga$. We shall find later 
in Sect. III that the analogous argument for the broken Weyl theory 
yields a constraint which is required to be satisfied for the
divergence relation following from the $g_{\mn}$-equations --- representing
the over-all energy-momentum conservation in the broken case --- to
vanish again.

Let us now turn to the conservation relations for the currents defined
in Eqs. (\ref{241}) -- (\ref{244}) as well as for the total
electromagnetic current. Introducing the charge operator
\be
\q = \ha(1 + \tau_3)
\label{262}
\ee
for the $\phi$-field the total electromagnetic current is given by
\begin{eqnarray}
j^{(e.m.)}_\r ={1 \over {K\sqrt{-g}}}\left({1\over\gtt}J_\r + 
{1\over\gt}J_{3\r}\right)
&\equiv& j^{(\phi)}_\r+j^{(\pl)}_\r+j^{(\pr)}_\r +j^{(\phi)}_{3\r}+j^{(\pl)}_{3\r}
\nonumber\\
&=&{i\over 2}\left(\phi^\da \q\rappD_\r\phi - \phi^\da\lappD_\r \q\phi\right)
-\bar\psi_e\ga_\r\psi_e ~,
\label{263}
\end{eqnarray}
Here $\q$ projects out the $\vphi_+$-component of $\phi$ with charge
$q=+1e$ in the $\phi$-part of the electromagnetic current
while the last term in (\ref{263}) yields, with
$-(\bar e_L\ga_\r e_L)-(\bar e_R\ga_\r e_R)=-\bar\psi_e\ga_\r\psi_e$, the
electromagnetic current contribution of the charged lepton (electron) with
$q=-1e$.

It is easy to show from the field equations (\ref{235}) and (\ref{236}) and
their adjoints that the Weyl-covariant divergence of the currents (\ref{242})
yield
\be
D^\r j^{(\pr)}_\r = -2D^\r j^{(\pl)}_\r = i\tilde\ga
\left[(\bar\pl \phi )\pr - \bar\pr(\phi^\da \pr)\right]\/.
\label{264}
\ee
This implies for the fermion part of the electromagnetic current the relation
\be
\appD\!{}^\r\left[j^{(\pl)}_\r + j^{(\pr)}_\r + j^{(\pl)}_{3\r}\right] =
i\tilde\ga\left[(\bar\pl \q\phi)\pr - \bar\pr(\phi^\da \q\pl)\right]\/,
\label{265}
\ee
where we have, moreover, used (for $a=3$) the divergence relation for the
fermionic isospin current
\be
\appD\!{}^\r j^{(\pl)}_{a\r} \equiv D^\r j^{(\pl)}_{a\r}-
\tilde g \ve_{abc}A^b_\r j^{(\pl)\r}_c
= i\tilde\ga\left[(\bar\pl\ha\tau_a\phi)\pr -
                              \bar\pr (\phi^\da \ha\tau_a\pl)\right]
\label{266}
\ee
following from Eqs. (\ref{244}) and (\ref{235}).
On the other hand, one concludes from Eqs. (\ref{238}) and (\ref{239})
that the hypercharge and isospin currents appearing on the r.-h. sides
of these equations are covariantly conserved.

Similarly, one concludes from the field equation (\ref{234}) and
its adjoint that the $\phi$-part of the isospin current obeys
\be
\appD\!{}^\r j^{(\phi)}_{a\r} = -i\tilde\ga\left[(\bar\pl\ha\tau_a\phi)\pr -
                              \bar\pr(\phi^\da\ha\tau_a\pl)\right]\/,
\label{267}
\ee
yielding for the sum $j^{(\phi)}_\r+j^{(\phi)}_{3\r}$ the relation
\be
\appD\!{}^\r\left[j^{(\phi)}_\r + j^{(\phi)}_{3\r}\right] =
-i\tilde\ga\left[(\bar\pl \q\phi)\pr - \bar\pr(\phi^\da \q\pl )\right]\/.
\label{268}
\ee
Eqs. (\ref{265}) and (\ref{268}) together, finally, lead for the total
electromagnetic current (\ref{263}) to the conservation relation
\be
D^\r j^{(e.m.)}_\r = 0\/,
\label{269}
\ee
where, for a fixed $3$-direction, we have replaced the $SU(2)_W$
and Weyl-covariant derivative $\appD_\r$ by $D_\r$.

As usual we now introduce the Weinberg angle $\tw$ by the rotation
relating $A_{3\r}$ and $B_\r$ to the $q=0$ component
of the the $SU(2)_W$ gauge fields $Z_\r$ and the electromagnetic fields
$A_\r$:
\begin{eqnarray}
A_{3\r} &=&~~ \cos\tw ~Z_\r + \sin\tw ~{e\over \hbar c}A_\r
\nonumber\\
B_\r &=& -\sin\tw ~Z_\r + \cos\tw ~{e\over \hbar c}A_\r \/.
\label{270}
\end{eqnarray}
The factor $e/\hbar c$ in front of the photon field
is introduced here for dimensional reasons
with the fields $A_{3\r}, B_\r , Z_\r ,$ and $\eh A_\r$ all
having length dimension $[L^{-1}]$.
Below we shall frequently abbreviate the electromagnetic potential
$\eh A_\mu$ with dimension $[L^{-1}]$ as $\bar A_\mu$. The
dimensionless gauge coupling constants $\gt$ and $\gtt$ were introduced
in Eq. (\ref{227}). In addition we shall introduce below the
dimensionless coupling constant $\gz$ for the neutral fields $Z_\r$
and a dimensionless strength for the electromagnetic coupling
given by $\e =\sqrt{4\pi\a_F}$ with $\a_F =e^2/\hbar c = 1/137.04$
denoting the fine-structure constant. As usual the elementary
electromagnetic charge --- a quantity with a dimension --- is denoted by 
$e$ with $-1|e|=-4.8032\cdot 10^{-10}\,esu$
being the electron charge. [$1\,esu = 1\,dyn^{\ha}\,cm$; we are using
$cgs$-units as conventional reference units. The intrinsic length
unit obtained after Weyl-symmetry breaking will be introduced
in Sect. III below.]

Using now a spherical basis for the isovector contributions
in Eq. (\ref{261a}) introducing the following charge changing current
components with $\Delta q=\pm 1$ [compare (\ref{261})]
\be
J^{(1)}_\r = {1 \over \sqrt{2}}\left(J_{1\r} - iJ_{2\r}\right)\/;\qquad
J^{(1)}_\r {}^\da = {1 \over \sqrt{2}}\left(J_{1\r} + iJ_{2\r}\right)\/,
\label{273}
\ee
and the corresponding gauge fields
\be
W_\r = {1 \over \sqrt{2}}\left(A_{1\r} - iA_{2\r}\right)\/;\qquad
W_\r{}^\da = {1 \over \sqrt{2}}\left(A_{1\r} + iA_{2\r}\right)\/,
\label{274}
\ee
and expressing, furthermore, the fields $A_{3\r}$ and $B_\r$ in terms
of the fields $Z_\r$ and $A_\r$ according to (\ref{270}) the interaction
Lagrangean (\ref{261a}) of Weyl weight zero may be written as
\be
\LWo_{int}~=-K\sqrt{-g}\left(\e\, j^{(e.m.)}_\r \bar A^\r +
\gz\, j^{(0)}_\r Z^\r\right)
-\left(J^{(1)}_\r{}^\da W^\r + J^{(1)}_\r W^\r{}^\da\right)\/.
\label{275}
\ee
Here $j^{(e.m.)}_\r$ is the electromagnetic current defined in
(\ref{263}). In order to obtain the electromagnetic interaction
in the conventional form given in (\ref{275}) one has to demand
that the coupling constants $\gt$ and $\gtt$ and the Weinberg angle
$\tw$ are related by
\be
\gt\,\sin \tw = \gtt \cos \tw = {\gt\gtt\over{\sqrt{\gt^2+\gtt^2}}} = \e \/,
\label{276}
\ee
where $\e$ is the dimensionless electromagnetic coupling strength
introduced above. The second term in the first bracket of
(\ref{275}) describes the coupling of the neutral
weak current $j^{(0)}_\r$ to the neutral gauge fields $Z_\r$ with
a coupling strength given by 
\be
\tilde g_0 = {\gt\gtt\over{\e}} = {\gt\over{\cos\tw}} = \sqrt{\gt^2 + \gtt^2}\/,
\label{276a}
\ee
where $j^{(0)}_\r$ is defined by
\be
j^{(0)}_\r = j^{(\phi)}_{3\r} + j^{(\pl)}_{3\r} - \sin^2\tw ~j^{(e.m.)}_\r \/.
\label{277}
\ee
All these definitions are the same as those appearing in the usual
formulation of the standard electroweak model.

Along with $J^{(1)}_\r$ and $J^{(1)}_\r{}^\da$ one may now also introduce
the current densities $J^{(e.m.)}_\r$ and  $J^{(0)}_\r$ of
Weyl weight zero by
\be
J^{(e.m.)}_\r = K\sqrt{-g}~\e j^{(e.m.)}_\r ~;\qquad
J^{(0)}_\r = K\sqrt{-g}~\gz j^{(0)}_\r\/.
\label{278}
\ee

The last two terms in Eq. (\ref{275}) represent the
coupling of the charged $SU(2)_W$ gauge fields to the charge
changing weak currents describing the weak decays of particles
in phenomenological applications. For low momentum transfer
processes, $p^2\approx 0$, we have by identification with the
effective current-current theory of weak interactions with the
Fermi constant $G_F$ the relation
\be
\left(\ha\,{\gt\over{\sqrt{2}}}\right)^2\,{1\over{m_W^2}} =
{G_F\over{\sqrt{2}}}\/,
\label{278a}
\ee
where $m_W$ is the mass of the $W$-field defined in Eq. (\ref{347})
below. Compare in this context also Eq. (\ref{261}) as well
as (\ref{328}) below.\\

\noindent{\bf C. Remarks Concerning Symmetry Breaking}

We, finally, compute the square of the Dirac operators
\be
\Dtt = -i\ga^\mu\!\appD_\mu \quad {\rm and} \quad
\Dt = -i\ga^\mu\!\simD_\mu
\label{279}
\ee
in application to the fermion fields $\pl$ and $\pr$, respectively.
From the field equations (\ref{235}) and (\ref{236}) one easily
derives that
\be
\Dtt\Dtt\pl = -i\tilde\ga\ga^\mu\pr \appD_\mu\phi + \tilde\ga^2\phi(\phi^\da\pl)\/,
\label{280}
\ee
and\be
\Dt\Dt\pr = -i\tilde\ga (\phi^\da\lappD_\mu\ga^\mu\pl) +
\tilde\ga^2(\phi^\da\phi)\pr\/.
\label{281}
\ee
As regards the $SU(2)_W$ degrees of freedom Eq. (\ref{280}) is a
matrix equation for a two-component isospinor while (\ref{281}) is
an equation for an isoscalar with the round brackets denoting
$SU(2)_W$ invariants. The left- and right-handed fields are coupled
in these equations through the first terms on their r.-h. sides.
We observe that both equations {\it decouple} and become eigenvalue
equations for $\pl$ and $\pr$, respectively, which are moreover
{\it diagonal in spin space} provided the $\phi$-field is covariant
constant, i.e. obeys
\be
\appD_\mu \phi = 0\/.
\label{282}
\ee
A property of this type would annihilate the first term in the
field equations (\ref{235}) for $\p$ and would yield an algebraic
constraint involving the curvature scalar $R$, the $\p$-field, and
the $\psi$-fields. It is not immediately clear whether this is
consistent with the other field equations. Therefore we shall not
demand Eq. (\ref{282}) to be satisfied. However, an equation of a
similar nature will be used in Sect. III below when we investigate
the breaking of the Weyl symmetry to obtain, finally, a gauge
theory formulated in a Riemannian space-time in the limit.

After these remarks concerning the standard electroweak theory 
and its formulation in a Weyl geometric framework we turn, as mentioned,
to the breaking of the $G$-gauge symmetry. Let us, however, first
consider, at the end of this section, the so-called breaking of the
electroweak gauge symmetry $\simG = SU(2)_W\times U(1)_Y$ to the
electromagnetic gauge symmetry $\Uem$ which is described in the
standard model as the result of a so-called ``spontanous symmetry breaking
due to a nonvanishing vacuum expectation value of the scalar field''.

In order to view the situation more clearly we investigate
in Appendix A the coset representation for the field $\p$
in terms of transformations $U(\gp)$ or $U(\p)$ parametrizing
$\simG\!/H$ [see (A5) and (A11)] by
generating the function $\p$ from the real function
$\ph =\left({0\atop\hat\vphi_0}\right)$ with $\ph$
being invariant under the subgroup $H$ of $\simG$, where $H$ is
identified with the {\it electromagnetic} gauge group $\Uem = U(1)_+$ generated
by $\q$ and obeying $\q\,\ph =0$ [compare (A3) and (A7)]. The transition
from $\p$ to $\ph$ with the help of the transformation $U^{-1}(\p)$
is to be regarded as a {\it choice of coordinates} for the 
representation of the scalar field in the theory and has,
in the first place, nothing to do with a ``vacuum expectation
value'' of this field. To adopt the origin $\ph$ in $\simG\!/H$
as a parametrization for the scalar field is done for physical
reasons establishing thereby the electromagnetic gauge group
$H = \Uem$ as stability group of the point $\ph$ in the
formalism and relate it to physical observations and experiments.
This choice is, actually, not a breaking of the original $\simG$-gauge theory
but a different realization of it. It is thus better to say
that after transforming $\p$ to $\ph$ with the help of $U^{-1}(\p)$
one has adopted an {\it electromagnetic gauge} in the electroweak theory
with the residual gauge transformations being given by $U(h(\a))\in\Uem$
[see (A7), (A15)  and the relation of these transformations to the so-called
``Wigner rotations'']. A true symmetry reduction from $G$ to a
subgroup of $G$ is governed by a relation of the type (\ref{282}) to
which we turn in the next section \cite{5}.

Conditions of the type of Eq. (\ref{282}) where $\p$ is a section
on a bundle with a homogeneous space as fiber are well-known from
differential geometry. They guarantee that in a reduction of the
structural group of a certain principal bundle $P(B,G)$ over the
base $B$ and with structural group $G$ to a bundle $P'(B,G')$
over the same base and with a subgroup $G'$ of $G$ as structural
group also the corresponding {\it connection} reduces from a
${\bf g}$-valued to a ${\bf g'}$-valued form, where ${\bf g}$ and
${\bf g'}$ denote the respective Lie algebras of the groups $G$
and $G'\subset G$ (compare \cite{6} for details). In concluding this
section let us, therefore, state the following theorem well-known from
the literature on differential geometry: {\it As the condition for
a true symmetry reduction in the physical sense from a theory with
gauge group $G$ to a theory with gauge group $G'\subset G$, implying
also the reduction of the connection on $P(B,G)$ to the connection
on $P'(B,G')$, it is required that there exists a section $\p_E$ 
on the bundle $E(B,G/G',G)$, associated to $P(B,G)$, with fiber $F=G/G'$ and
structural group $G$ which is covariant
constant, i.e. obeys $D\p_E = 0$, where $D$ is the covariant
derivative on $E$.}

\section{Weyl-Symmetry Breaking}
As a term in the Lagrangean which breaks the Weyl-symmetry with the aim
of introducing a scale of lengths with the help of the $\p$-field and
establish an electroweak theory of leptons in the presence
of gravitation formulated in a $V_4$, we add the following expression
of Weyl weight $+1$ to the Weyl-invariant Lagrangean density
$\LW$ given in (\ref{228}):
\be
\L_B = -{a\over 2}K\sqrt{-g}\left\{{1\over 6}R + \left[{mc\over\hbar}\right]^2
        \p^\da\p\right\}\/.
\label{31}
\ee
Here $a$ is a dimensionless constant, $R$ is the curvature scalar of
the ambient Weyl space $W_4$ [see (\ref{230})] which is related here to the
mass --- or rather Compton wave length --- of the universal scalar field $\p$
by tying $R$ to the squared modulus $\Phi^2 = \p^\da\p$ of this field.
The expression (\ref{31}) is independent of the standard model gauge fields
associated with the group $\simG$ and thus leaves the $\simG$-gauge invariance
unaffected. However, the explicit breaking of the $D(1)$ symmetry caused
by $\L_B$ will lead, as we shall see, to nonzero masses not only for the
$\p$-field but also for the fermion and the gauge boson fields in a manner
similar to the situation realized in the standard electroweak theory.

We have mentioned at the end of Sect. II that the standard model is not
characterized by a true symmetry reduction from a gauge group $\simG$
to a subgroup of $\simG$. What is conventionally called a spontanous
symmetry breaking is a choice of an appropriate coordinatization
taking due recognition of the electromagnetic phase group $\Uem$
as a subgroup generated by $\q$ in the formalism. On the contrary, 
adding (\ref{31}) to the Lagrangean $\LW$ will lead to a true
breaking of the $G$-gauge theory [see (\ref{215})] to a theory
with the subgroup $G\,'=SO(3,1)\otimes\simG$ of $G$ as gauge group.
It is the square of the modulus, $\Phi^2$, of the $\p$-field which
is the section on $E(W_4,G/G\,',G)$, required to be covariant constant
in the sense of the theorem quoted at the end of Sect. II, which
governs, as we shall see, the symmetry breaking by (\ref{31})
yielding, ultimately, a $V_4$ from a $W_4$ and the generation of
nonvanishing masses. Hence the symmetry breaking relation will be
\be
D_\mu \Phi^2 \equiv \partial_\mu\Phi^2 + \ka_\mu\Phi^2 = 0\/,
\label{32}
\ee
with $\Phi^2\in G/G\,'\equiv D(1)$.

However, before we come to this point, let us first derive the field
equations following from a variational principle formulated with the
Lagrangean $\L$ given by
\be
\L = \LW +~ \L_B \/.
\label{33}
\ee
One finds using the same notation as above:
\begin{eqnarray}
&&\d\p^\da :~~g^{\mn}\appD_\mu\appD_\nu\p + {1\over 6}R\p + 4\beta
  (\p^\da\p)\p - 2\tilde\ga\bar\pr\pl + a\left[{mc\over\hbar}\right]^2\p = 0\/,
  \label{34}\\
&&\d\pl^\da :~~-i\ga^\mu\appD_\mu\!\pl - \tilde\ga\p\pr = 0\/,\label{35}\\
&&\d\pr^\da :~~-i\ga^\mu\simD_\mu\!\pr - \tilde\ga (\p^\da\pl) = 0\/,\label{36}\\
&&\d\ka_\r :~~\tilde\d D_\mu f^{\mu\r} = -6\tilde\a D^\r R + {a\over 4}\ka^\r\/,
   \label{37}\\
&&\d B_\r :~~\dtt D_\mu B^{\mu\r} = \gtt\left[j^{(\p)\r} + j^{(\pl)\r} +
              j^{(\pr)\r}\right]\/,\label{38}\\
&&\d A^a_\r :~~\dtt \appD_\mu\!F^{\mu\r}_a = \gt\left[j^{(\p)\r}_a +
              j^{(\pl)\r}_a\right]\/, \label{39}\\
&&\d g_{\mn} :~~{1\over 6}(\Phi^2+a)\left[R_{(\mn)}-\ha g_{\mn}R\right]-
   4\tilde\a R\left[R_{(\mn)}-{1\over 4}g_{\mn}R\right]-4\tilde\a\left\{
   D_{(\mu}D_{\nu)}R-g_{\mn}D^\r D_\r R\right\} = 
   \nonumber\\
&&\qquad\qquad =\Theta^{(\p)}_{\mn} + T^{(\pl)}_{\mn} + T^{(\pr)}_{\mn} +
   T^{(f)}_{\mn} + T^{(B)}_{\mn} + T^{(F_a)}_{\mn} - g_{\mn}\tilde\ga
   \left[(\bar\pl\p)\pr+(\bar\pr(\p^\da\pl)\right] + \nonumber\\
&&\qquad\qquad\qquad\qquad +\,g_{\mn}{a\over 2}\left[{mc\over\hbar}\right]^2\Phi^2\/,
   \label{310}\\
&&\d a :~~{1\over 6}R + \left[{mc\over\hbar}\right]^2\Phi^2 = 0\/,
\label{311}
\end{eqnarray}
where Eqs. (\ref{35}), (\ref{36}), (\ref{38}) and (\ref{39}) are unchanged;
compare Eqs. (\ref{235}), (\ref{236}), (\ref{238}) and (\ref{239}).
The energy-momentum tensors appearing in (\ref{310}) are the same as those
defined in Eqs. (\ref{245}) -- (\ref{249}) of Sect. II. Only the
energy-momentum tensor for the scalar field is to be changed now, for
$a \neq 0$, to the expression given by the sum of the first and the last
term on the r.-h. side of (\ref{310}), i.e. by
\be
\Theta^{(\p)}_{\mn}\,' = \Theta^{(\p)}_{\mn} + 
              g_{\mn}\,{a\over 2}\left[{mc\over\hbar}\right]^2\Phi^2\/.
\label{311a}
\ee

We first turn to the trace condition following from (\ref{310}). The trace
of $\Theta^{(\p)}_{\mn}\,'$ for solutions of the field equation (\ref{34})
and its adjoint is now given by
\be
\Theta_\mu^{(\p)}\,'{}^\mu = -{1\over 6}R\p^\da\p + \tilde\ga\left[(\bar\pl\p)\pr +
       \bar\pr(\p^\da\pl)\right] + a\left[{mc\over\hbar}\right]^2\p^\da\p\/,
\label{312}
\ee
while the other traces of the energy-momentum tensors are the same as
in Sect. II. With these and Eq. (\ref{312}) one concludes from the trace
of (\ref{310}) that
\be
\tilde\a D^\r D_\r R = 0\/.
\label{313}
\ee
Taking the Weyl-covariant divergence of Eq. (\ref{37}) one finds with 
(\ref{313}) that the Weyl vector fields must satisfy the Lorentz-like
condition:
\be
D_\r \ka^\r \equiv \bar\nabla_\r \ka^\r - \ka_\r\ka^\r = 0 \qquad {\rm for}
      \qquad a\neq 0\/,
\label{314}
\ee
where $\bar\nabla_\r$ denotes the metric covariant derivative
[compare Appendix A of I]. Eq. (\ref{314}) in turn implies that the
$W_4$ curvature scalar (\ref{230}) is now given by
\be
R = \bar R - {3\over 2}\ka_\r\ka^\r\/.
\label{315}
\ee

We, finally, compute the divergence conditions for the solutions of the
field equations (\ref{34}) -- (\ref{311}) which follow from (\ref{310})
by taking the Weyl-covariant divergence of this equation and using the
contracted Bianchi identities (I, A40) for the $W_4$ as well as the
equations
\be
D^\mu T^{(f)}_{\mn} = f_\nu{}^\r \left[\,6\tilde\a D_\r R - {a\over 4}\ka_\r\right]\/.
\label{316}
\ee
Before the symmetry breaking by $\L_B$ these energy-momentum balance
relations for the set of interacting fields were identically satisfied in Sect. II.
In the broken case we now obtain that the following relations must hold
for the divergence relations, deduced from (\ref{310}), to be fulfilled
again [compare (I, 4.16)]~:
\be
D^\mu f_{\mn} - 3f_{\nu\mu}\ka^\mu = 0 \qquad {\rm for} \qquad a\neq 0\/.
\label{317}
\ee
These relations are trivially satisfied for a Weyl vector field being
``pure gauge'', i.e. implying $f_{\mn} = 0$. This is identical with the
condition (\ref{32}) being satisfied, which may be written as
\be
\ka_\mu = -\partial_\mu\log\Phi^2\/.
\label{318}
\ee
The Weyl vector field in this broken Weyl theory is thus derivable
from a potential given by the modulus of the scalar field. This is
in direct analogy to the case of the Christoffel connection,
$\bar\Gamma_{\mn}{}^\r=\{{\r\atop\mn}\}$, following from the
relation $\bar\nabla_\r g_{\mn} = 0$ in (pseudo-)Riemannian geometry
with $g_{\mn}\in Gl(4,R)/SO(3,1)$.

The field equations (\ref{37}), finally, lead --- together with
(\ref{311}) which implies $D_\r R = 0$ for $D_\r \Phi^2 = 0$ --- to
\be
\ka_\mu = 0\/;\qquad {\rm i.e.}\qquad \Phi^2 = const \/.
\label{319}
\ee
This shows that the Weyl space $W_4$ reduces completely to a
pseudo-Riemannian space $V_4$ with the scalar field possessing a
constant modulus. The value of this modulus can not be computed
numerically. On the other hand, the value for $\Phi$ will determine
the fermion and gauge boson masses appearing in the broken Weyl theory
as well as in the nonlinearity contained in the $\p$-equation
yielding the ``Higgs dynamics'' in the standard model. A fixing of
an ungauged $D(1)$ degree of freedom is, indeed, implicit in the
standard model. However, a relation of the type (\ref{32}) for
the $D(1)$ gauge symmetry breaking does not appear in the standard 
model since this would require to go beyond the flat space formulation
of the conventional description. In the present context we have to
investigate, in the $V_4$ limit given by (\ref{319}), the appearance
of Einstein's equations for the metric coupled to the energy-momentum
tensors of the now massive fermion and gauge boson fields and establish
the fact that gravitation, as we know it from general relativity, is
a natural part of the broken $G$-gauge dynamics described by $\L$.\\

\noindent{\bf A. Electromagnetism in the WEW Theory}

In this subsection we first turn to the field equations for the
electromagnetic fields $F_{\mn} = \pa_\mu A_\nu - \pa_\nu A_\mu$
and the fields $Z_{\mn} = \pa_\mu Z_\nu - \pa_\nu Z_\mu$ following
from Eqs. (\ref{38}) and (\ref{39}). The total $\simG$-curvature,
written in Lie algebra valued form with $\F_{\mn} = F^a_{\mn}\,\ha\tau_a$,
is [compare Eqs. (\ref{231}) and (\ref{232})]
\be
\F_{\mn} + \ha\,{\bf 1} B_{\mn} = \pa_\mu\!\Gt_\nu - \pa_\nu\!\Gt_\mu
             +\, i\gt \left[\Gt_\mu ,\Gt_\nu\right]\/,
\label{320}
\ee
with $\Gt_\mu$ as given by (\ref{B9}). The definitions of the curvature
components $F^a_{\mn}$ and $B_{\mn}$ imply that in a spherical basis
we have, with Eqs. (\ref{270}) and (\ref{276}), the relations
\begin{eqnarray}
F^3_{\mn} &=& \gt\gz^{-1}\,Z_{\mn} + \gtt\gz^{-1}{e\over{\hbar c}}\,F_{\mn}
             - i\gt \left(W^\da_\mu W_\nu - W^\da_\nu W_\mu\right)\/,
\label{321} \\
B_{\mn} &=& -\gtt\gz^{-1}\,Z_{\mn} + \gt\gz^{-1}{e\over{\hbar c}}\,F_{\mn}\/,
\label{322} \\
F^-_{\mn} &=& \left(F^+_{\mn}\right)^\da = \pa_\mu W_\nu - \pa_\nu W_\mu
           -i \left\{\gt\cos\tw\;[W_\mu Z_\nu - W_\nu Z_\mu]
    + \e\;{e\over{\hbar c}}[W_\mu A_\nu - W_\nu A_\mu ]\right\}\/,
\label{323}
\end{eqnarray}
with $F^{\pm}_{\mn}={1\over{\sqrt 2}}(F^1_{\mn} \pm iF^2_{\mn})$. In (\ref{320})
$F^3_{\mn}\,\ha\tau_3 +\ha\,{\bf 1}B_{\mn}$ is that part
which commutes with $\q$ while $F^{\pm}_{\mn}$, as defined by (\ref{323}),
denote the off diagonal parts which do not commute with $\q$.

We now rewrite the field equations (\ref{38}) and (\ref{39}) --- the latter
at first for $a=3$ --- using (\ref{321}), (\ref{322}) and the definition 
(\ref{263}) of the electromagnetic current and find , with $\appD_\mu
F_3^{\mu\r}\equiv D_\mu F_3^{\mu\r}$ and Eqs. (\ref{276}) and (\ref{276a}),
\be
{e\over{\hbar c}}\,D_\mu F^{\mu\r} = {\dtt}{}^{-1}\,\e\, j^{(e.m.)\r} + 
     i\,\e\,D_\mu \left(W^\da{}^\mu W^\r - W^\da{}^\r W^\mu\right)\/,
\label{324}
\ee
and, with $j^{(0)}_\r$ as defined in (\ref{277}),
\be
D_\mu Z^{\mu\r} = {\dtt}{}^{-1}\,\gz\,j^{(0)\r} + i\,\e\,\gt\gtt^{-1}\,D_\mu
                        \left(W^\da{}^\mu W^\r - W^\da{}^\r W^\mu\right)\/.
\label{325}
\ee
We continue to write here the covariant derivative as $D_\mu$ disregarding for
the moment that $\ka_\mu =0$ according to (\ref{319}) in the broken case
since (\ref{324}) and (\ref{325}) are valid also in the Weyl symmetric
theory discussed in Sect. II. The l.-h. side of (\ref{324}) could also be
written $D_\mu\bar F^{\mu\r}$ with the electromagnetic field strengths
$\bar F^{\mu\r}=\pa^\mu\bar A^\r - \pa^\r\bar A^\mu$ of dimension $[L^{-2}]$.
Eqs. (\ref{324}), moreover, show that besides the electromagnetic source
current $j^{(e.m.)}_\r$ there contributes also a term on the r.-h.\,side
which is bilinear in the $W_\mu$-fields with $W_\mu$ and $W^\da_\mu$ being
related to the charge changing weak processes. The same remark applies to the
last term in (\ref{325}) representing the contribution of the 
$W_\mu$-fields to the neutral weak processes.

Furthermore, Eqs. (\ref{324}) and (\ref{325}) imply current conservation
[compare (\ref{269})]
\be
D^\r j_\r^{(e.m.)}\equiv \bar\nabla^\r j^{(e.m.)}_\r = {1\over{\sqrt{-g}}}
             \pa_\mu\left(\sqrt{-g}\,g^{\mu\r}\,j^{(e.m.)}_\r\right) = 0 \/,
\label{325c}
\ee
and, analogously, $D^\r j_\r^{(0)}=0$
with the Weyl-covariant divergence of the $W$-terms and of the 
l.-h. sides of these equations yielding zero. Writing
Eqs. (\ref{324}) and (\ref{325}) in the electromagnetic gauge (see Appendix B)
it follows from (\ref{B5}) -- (\ref{B8}) and (\ref{B16}) that these
equations are $\Uem$ gauge invariant.

It may be worth while in this context to write down explicitly the $\Uem$ gauge
invariant source currents $\hat j^{(e.m.)}_\r$ and $\hat j^{(0)}_\r$ in the 
electromagnetic gauge [compare Eqs. (\ref{263}), (\ref{277}), (\ref{B3}),
(\ref{B14}) and (\ref{B15})]
\begin{eqnarray}
\hat j^{(e.m.)}_\r &=& - \hat{\bar e}_L\ga_\r\hat e_L - \hat{\bar e}_R\ga_\r\hat e_R
\nonumber \\
  &=& - (\hph)^{-2}\Biggl\{|\vv_+|^2\left(\bar\nu_L\ga_\r\nu_L\right) -
    |\vv_+|^2\left(\bar e_L\ga_\r e_L\right) + \vv_+\vv^*_0\left(\bar\nu_L\ga_\r e_L\right)
\nonumber \\
&\hfil& \qquad +\vv^*_+\vv_0\left(\bar e_L\ga_\r\nu_L\right)\Biggr\} -
                              \bar\psi_e\ga_\r\psi_e\/,
\label{324a}
\end{eqnarray}
and
\be
\hat j^{(0)}_\r = -{1\over 8}\gt\gtt (\hph)^2\,\left(\hat Z_\r 
        + \hat Z_\r^\da\right) + \ha\left(\hat{\bar\nu}_L
     \ga_\r\hat\nu_L\right) - \ha\left(\hat{\bar e}_L\ga_\r\hat e_L\right) -
         \sin^2\tw\;\hat j^{(e.m.)}_\r \,.
\label{325a}
\ee

We, finally, determine the field equations for $F^{\pm}_{\mn}$, i.e. for
$a=1,2$ in (\ref{39}). They read:
\begin{eqnarray}
\appD_\mu (F^-){}^{\mu\r} &\equiv& D_\mu(F^-){}^{\mu\r} - i\gt\, W_\mu F^{\mu\r}_3
  +i\gt\, A^3_\mu(F^-){}^{\mu\r} =\, {\dtt}{}^{-1} \gt\; j^{(1)\r} \/,
\label{326} \\
\appD_\mu (F^+){}^{\mu\r} &\equiv& D_\mu (F^+){}^{\mu\r} +i\gt\,W^\da_\mu F^{\mu\r}_3
  -i\gt\, A^3_\mu(F^+){}^{\mu\r} =\, {\dtt}{}^{-1} \gt\; j^{(1)\r\da}\/,
\label{327}
\end{eqnarray}
where [compare (\ref{261}) and (\ref{273})]
\be
j^{(1)}_\r = {1\over{\sqrt 2}}\left(j^{(\p)}_{1\r} + j^{(\pl)}_{1\r} -
              i\,j^{(\p)}_{2\r} - i\,j^{(\pl)}_{2\r}\right)
\label{328}
\ee
is the charge changing current. In (\ref{326}) and (\ref{327}) $F_3^{\mu\r}$ and
$A^3_\mu$ may be replaced according to Eqs. (\ref{321}) and (\ref{270}) in
order to yield field equations involving only the fields $A_\mu, Z_\mu$ and
$W_\mu , W^\da_\mu$. Written in the electromagnetic gauge Eqs. (\ref{326})
and (\ref{327}) are again $\Uem$ gauge covariant. To establish this result
one needs the formula
\be
\left(\hat F^{\mp}_{\mn}\right)' = e^{\mp i{e\over{\hbar c}}\a(\p ',\p)}\;
                                 \hat F^{\mp}_{\mn}\/,
\label{329}
\ee
which is easily derivable from (\ref{323}) with the help of
(\ref{B5}) -- (\ref{B8}). Moreover, one needs the following expression
for the the current $\hat j^{(1)}_\r$ evaluated at the origin in $\simG\!/H$.
With the help of Eqs. (\ref{B3}) and (\ref{B15}) one finds
\be
\hat j^{(1)}_\r = - {1\over 4}\gt\, (\hph)^2\,\hat W_\r + {1\over{\sqrt 2}}\,
                \hat{\bar e}_L\ga_\r\hat\nu_L \/.
\label{330}
\ee
From the form of (\ref{330}) the transformation rule
\be
\left(\hat j^{(1)}_\r\right)' = e^{-i{e\over{\hbar c}}\a(\p ',\p)}\,\hat j^{(1)}_\r
\label{331}
\ee
under residual $\Uem$ gauge transformations is at once apparent as a
consequence of (\ref{B6}) and (\ref{B16}), and correspondingly for
$\hat j_\r^{(1)}{}^\da$ appearing in Eq. (\ref{327}) after transformation
to the electromagnetic gauge.

In order to gain information about the constant $\dtt$ in
Eqs. (\ref{324}) and (\ref{325}) and bring (\ref{324}) 
---  disregarding the $W_\mu$-contributions for a moment --- into the
form of Maxwell's equations in
electromagnetism, we first observe that each term in these
equations has length dimension $[L^{-3}]$. Multiplying (\ref{324}) by the
charge $e$ and introducing the fine-structure constant 
$\a_F=\e^2/4\pi$ we can rewrite Eqs. (\ref{324}) and (\ref{325})
as
\begin{eqnarray}
D_\mu F^{\mu\r} &=& {4\pi\over{c}}\,{\bf j}^{(e.m.)\r} +i\,{4\pi\over\e}\,
             e\,D_\mu\left(W^{\da\mu}W^\r - W^{\da\r}W^\mu\right)\/,
\label{334} \\
D_\mu Z^{\mu\r} &=& {1\over{c}}\,{\bf j}^{(0)\r} + i\,\e\,\gt\gtt^{-1}
                    D_\mu\left(W^{\da\mu}W^\r - W^{\da\r}W^\mu\right)\/,
\label{335}
\end{eqnarray}
by taking (compare (\ref{346a}) and (\ref{346b}) below)
\be
\dtt = \dtt{}\!'\,\lp^2  \qquad \mbox{with} \qquad \dtt{}\!'= {1\over\e} \/,
\label{336}
\ee
Furthermore, we have introduced here the following current densities:
\be
{\bf j}^{(e.m.)\r} = e\,c\,\lp^{-2}\,j^{(e.m.)\r}\/;\qquad
{\bf j}^{(0)\r} = \e \gz\,c\,\lp^{-2}\,j^{(0)\r}\/.
\label{337}
\ee

We have measured in (\ref{336}) the constant $\dtt$ of dimension
$[L^2]$ in units of $\lp^2$ defined in Eq. (\ref{346a}) and
determined the numerical coefficient $\dtt{}\!'$ in such a way that
the first term on the r.-h. side of (\ref{334}) has the
conventional form known from electromagnetism. This fixed the
numerical constant $\dtt{}\!'$ to the value $\e^{-1}=1/\sqrt{4\pi\a_F}$.\\

\noindent{\bf B. Gravitation in the Broken WEW Theory}

Einstein's equations for the metric follow from (\ref{310}), as we
will now show, with a total energy-momentum tensor $T_{\mn}$ on
the r.-h. side for all the interacting massive and massless
fields involved. Clearly, $T_{\mn}^{(f)}=0$ as a consequence
of (\ref{319}). We first turn to the contributions of the
$\simG$-gauge fields contained in $T_{\mn}^{(B)}+T_{\mn}^{(Fa)}$
on the r.-h. side of (\ref{310}) and split this expression
into the familiar electromagnetic contribution, a $Z_\mu$-contribution,
and $W_\mu$-contributions in the following way:
\be
T_{\mn}^{(B)} + T_{\mn}^{(Fa)} = T_{\mn}^{(F)} + T_{\mn}^{(Z)} +
                                 T_{\mn}^{(W)} + T_{\mn}^{(WW)} \/.
\label{340}
\ee
Using (\ref{336}) we have here introduced the following
energy-momentum tensors possessing the dimension $[L^{-2}]$:
\be
T_{\mn}^{(F)} = - {\e\over{4\pi}}\,{1\over{\hbar c}}\,\lp^2\,
                 \left[F_{\mu\s}F_\nu{}^\s
                -{1\over 4}\,g_{\mn}F^{\r\l}F_{\r\l}\right]
\label{341}
\ee
for the electromagnetic fields, and
\begin{eqnarray}
T_{\mn}^{(Z)} &=& -{1\over\e}\,\lp^2\,\left[Z_{\mu\s}Z_\nu{}^\s
                  -{1\over 4}\,g_{\mn}\,Z^{\r\l}Z_{\r\l}\right]\/,
\label{342} \\
T_{\mn}^{(W)} &=& -{1\over\e}\,\lp^2\,\left[F^+_{\mu\s}F_\nu^{-\s}
    + F^-_{\mu\s}F_\nu^{+\s} - \ha\,g_{\mn}\,F^{+\r\l}F^-_{\r\l}\right]\/,
\label{343}
\end{eqnarray}
for the $Z$- and $W$-fields, respectively. The last term, $T_{\mn}^{(WW)}$,
in (\ref{340}) is a lengthy expression constructed with 
the fields $Z_{\mn}, F_{\mn}$
and $W^\da_\mu ,W_\nu$ containing terms of second and fourth order in the
$W$-fields which we shall not write down explicitly. All terms in
(\ref{340}) are traceless, so that the contributions of the masses
for the $Z$- and $W$-fields cannot come from these expressions but
must be contained in the other contributions on the r.-h side of (\ref{310}).
In fact, it is the energy-momentum tensor of the $\p$-field, (\ref{311a}),
which contains --- besides the mass $\sqrt{a}\,m$ of the $\p$-field
itself --- the effects of the nonzero masses for the $Z$- and
$W$-fields. In order to see this more clearly we consider the
symmetry breaking relation (\ref{32}) which implies, by taking the
Weyl-covariant divergence,
\be
D^\mu D_\mu \Phi^2 = 0\/.
\label{344}
\ee
Relating this to the field equation (\ref{34}) for $\p$ and
its adjoint leads to the following result:
\be
\left(\appD_\mu\!\p\right){}\!^\da\left(\appD{}\!^\mu\p\right) =
{1\over 6}R\p^\da\p
+ 4\b\left(\p^\da\p\right)^2 + a\left[{mc\over\hbar}\right]^2 \p^\da\p
      - \tilde\ga\left[(\bar\pl\p)\pr + \bar\pr(\p^\da\pl)\right]\/.
\label{345}
\ee
Evaluating (\ref{345}) at the origin $\ph$ of $\simG\!/H$ [compare Appendix A],
i.e. considering (\ref{345}) in the electromagnetic gauge, using moreover
$D_\mu\hph =0$ for $\hph\neq 0$, i.e. $\ka_\mu =0,~ \hph =const$ according
to (\ref{319}), one finds with the help of Eqs. (B3), (B18), (\ref{311}) 
and (\ref{315})
\begin{eqnarray}
\ha\gt^2\hph^2\,\hat W^\da_\mu \hat W^\mu + {1\over 4}\gz^2\,\hph^2\,
    \hat Z_\mu \hat Z^\mu &=& {1\over 6}\bar R\hph^2 + 4\b\hph^4 +
      a\left[{mc\over\hbar}\right]^2\hph^2 -
      \tilde\ga\hph\hat{\bar\psi}_e\hat\psi_e \nonumber \\
&=& 4\b\hph^4 + \left[{mc\over\hbar}\right]^2 \left(a-\hph^2\right)\hph^2
              - \tilde\ga\,\hph\,\hat{\bar\psi}_e\hat\psi_e \/.
\label{346}
\end{eqnarray}
This is an interesting relation between the mass terms for the various
fields involved in the theeory and the nonlinear self-coupling
term of the scalar field. 

We first observe that each term in (\ref{346}) has dimension $[L^{-2}]$
and that the unit of lengths in which every quantity with a length
dimension is to be measured is given by the length
\be
\lp = {\hbar\over{mc}}
\label{346a}
\ee
appearing on the r.-h. side of (\ref{346}). This length was introduced
by the Weyl-symmetry breaking Lagrangean $\L_B$ defined in (\ref{31}).
Indeed, for $a = 1$ the mass of the scalar field is $m_\p \equiv m$
and the corresponding Compton wave length is $\lp$. This length will from
now on be adopted as the unit of lengths. This choice implies that
all quantities with a length dimension have to be measured in units
of $\lp$ as we already did for the constant $\dtt$ in (\ref{336}).
For the constants $\b$ and $\tilde\ga$ this means that
\be
\b = \b '\,\lp^{-2}\/,\qquad \tilde\ga = - \tilde\ga '\,\lp^{-1}\/,
\label{346b}
\ee
with the primed quantities being numerical constants. The minus sign
in the second equation is adopted here in order to obtain the correct
sign for the electron mass in the last term of (\ref{346}) which
is thus given by
\be
- \tilde\ga\,\hph = \tilde\ga '\,\hph\,{mc\over\hbar} = {m_ec\over\hbar}\/.
\label{346c}
\ee
The same argument applies to the $Z$- and $W$-boson fields
of length dimension $[L^{-1}]$ and the corresponding masses.
We thus identify the masses of the charged ($q=\mp 1e$) and neutral 
($q=0$) boson fields as well as the electron mass $m_e$ by the equations:
\be
2\,m_W^2 = \ha\gt^2\,\hph^2\,m^2~;\qquad m_Z^2 = {1\over 4}\gz^2\,\hph^2\,m^2~;
\qquad m_e = \tilde\ga '\,\hph\,m~,
\label{347}
\ee
implying the relation
\be
m_Z = {m_W\over{\cos\tw}}
\label{348}
\ee
between the $Z$- and the $W$-masses, which is well-known from the
standard model. Below we shall sometimes denote by $\tilde m_W$,
$\tilde m_Z$ and $\tilde m_e$ the dimensionless quantities
\be
\tilde m_W = {m_W\over m} = \ha\gt\,\hph~;\qquad
\tilde m_Z = {m_Z\over m} = \ha\gz\,\hph~;\qquad \tilde m_e 
= {m_e\over m} = \tilde\ga '\,\hph \/.
\label{348a}
\ee

We now focus the attention on the energy-momentum tensor
$\Theta_{\mn}^{(\p)}{}'$ for the field $\p$ which was defined in
(\ref{311a}), reading with $a = 1$
\begin{eqnarray}
\Theta_{\mn}^{(\p)}{}' &=& \ha\,\left[\left(\appD_\mu\!\p\right)^\da
 \left(\appD_\nu\!\p\right) + \left(\appD_\nu\!\p\right)^\da
 \left(\appD_\mu\!\p\right)\right]
\nonumber \\
&\hfil& \qquad\qquad
- g_{\mn}\,\Biggl\{\ha\left(
 \appD\!{}^\r\p\right)^\da\left(\appD_\r\!\p\right)
- \b\,\left(\p^\da\p\right)^2
 - {1\over 2}\,\left[{mc\over\hbar}\right]^2\,\p^\da\p\Biggr\}\/.
\label{349}
\end{eqnarray}
Using now the relation (\ref{345}) for $a=1$ in order to to eliminate
the $(\p^\da\p)^2$-coupling term proportional to $\b$ in (\ref{349}) 
one obtains, considering also (\ref{310}) again,
\begin{eqnarray}
\Theta_{\mn}^{(\p)}{}' &=& \ha\,\left[\left(\appD_\mu\!\p\right)^\da
\left(\appD_\nu\!\p\right) + \left(\appD_\nu\!\p\right)^\da
\left(\appD_\mu\!\p\right)\right]
\nonumber \\
&\hfil& \quad
- {1\over 4} g_{\mn}\,\Biggl\{
\left(\appD\!{}^\r\p\right)^\da\left(\appD_\r\p\right) -
\left[{mc\over\hbar}\right]^2\,\left(\Phi^2 + 1\right)\Phi^2
-\tilde\ga\,\left[(\pl\p)\pr + \bar\pr(\p^\da\pl)\right]\Biggr\}\/.
\label{350}
\end{eqnarray}
Evaluating this in the electromagnetic gauge yields with Eqs. (B3),
(\ref{346c}), (\ref{347}), and (\ref{348a}), and with
$\hat Z_\mu^\da = \hat Z_\mu$, the result
\begin{eqnarray}
\hat\Theta_{\mn}^{(\p)}{}' &=& \tilde m_W^2\left(
\hat W_\mu^\da \hat W_\nu + \hat W_\nu^\da \hat W_\mu\right) + 
\tilde m_Z^2\,\hat Z_\mu \hat Z_\nu
\nonumber \\
&\hfil&\qquad
- {1\over 4} g_{\mn}\,\Biggl\{
2 \tilde m_W^2\,\hat W_\r^\da \hat W^\r + \tilde m_Z^2\,
\hat Z_\r \hat Z^\r
- \left[{mc\over\hbar}\right]^2\,\left(\hph^2 + 1\right)\hph^2 +
{m_ec\over\hbar}\,\hat{\bar\psi}_e\hat\psi_e\Biggr\}\/,
\label{351}
\end{eqnarray}
showing that the boson and fermion mass terms appear in the total
energy-momentum tensor through the tensor $\hat\Theta_{\mn}^{(\p)}{}'$.

The r.-h. side of (\ref{310}), i.e. the source term of the field
equations for the metric, when evaluated in the electromagnetic gauge
and for $\ka_\mu =0,~ \hph=const$ now reads
\be
\hat T_{\mn} = \Biggl\{\hat\Theta_{\mn}^{(\p)}{}' + \hat T_{\mn}^{(\pl)}
+ \hat T_{\mn}^{(\pr)} + g_{\mn}\,{m_ec\over\hbar}\,
\hat{\bar\psi}_e\hat\psi_e + \hat T_{\mn}^{(F)} + 
\hat T_{\mn}^{(Z)} + \hat T_{\mn}^{(W)} +
\hat T_{\mn}^{(WW)} \biggr\}_{\ka_\mu =0,~\hph =const} \/,
\label{352}
\ee
where $\hat\Theta_{\mn}^{(\p)}{}'$ is given by (\ref{351}), and (B18)
has been used for the Yukawa term. After Weyl-symmetry breaking
according to (\ref{31}), (\ref{32}) yielding (\ref{319})
we thus have to evaluate the r.-h. side of
(\ref{352}) for a vanishing Weyl vector field and for $\hph = const$,
i.e. in a $V_4$, which is indicated by the suffix on the curly
brackets. Moreover, we use the length $\lp$ as a universal unit.
By construction $\hat T_{\mn}$ satisfies the usual conservation
relations $\bar\nabla^\mu\hat T_{\mn} = 0$ due to
Eqs. (\ref{317}) -- (\ref{319}).

The l.-h side of (\ref{310}) for $R = \bar R = const$ according to
Eqs. (\ref{311}) and (\ref{319}), and with $\tilde\a =0$ --- remembering
that $\tilde\a$ was introduced in $\LW$, as discussed in I, to yield a
nontrivial dynamics for the Weyl vector fields which now vanish --- is given
in the electromagnetic gauge and in the $V_4$ limit, taking moreover $a=1$
(see above), by
\be
{1\over 6}\,\left(\hph^2 + 1 \right)\left[\bar R_{\mn}
               - \ha\,g_{\mn}\,\bar R \right] \/.
\label{353}
\ee
This together with (\ref{352}) yields, finally, a set of field equations
for the metric of the form
\be
\bar R_{\mn} - \ha\,g_{\mn}\,\bar R = {1 \over{{1\over 6}[\hph^2 + 1]K}}~
\hat T_{\mn}{}' \/.
\label{354}
\ee
Here appears the constant $K$ in the denominator on the
r.-h. side since we want to measure the total energy-momentum
tensor ultimately in the conventional units of $[Energy/L^3]$ while
$\hat T_{\mn}$ in (\ref{352}) has dimension $[L^{-2}]$. This we indicate
by a prime on the total source tensor which is given by
$\hat T_{\mn}{}' = K\,\hat T_{\mn}$.  Eqs. (\ref{354}) are
identical with Einstein's field equations in general relativity
provided we are entitled to make the dentification
\be
\ka_E = {1 \over{{1\over 6}[\hph^2 + 1]K}}~,
\label{355}
\ee
where $\ka_E$ is Einstein's gravitational constant, $\ka_E = 8\pi N/c^4 =
2.076\cdot 10^{-48} g^{-1} cm^{-1} sec^2$, and $N$ is Newton's
constant. Of course, in the framework adopted here also $\ka_E$
is to be expressed in the proper units related to $\lp$ as the chosen
intrinsic fundamental length unit replacing the $cgs$-units
conventionally chosen for $\ka_E$ yielding the quoted numerical value
of this constant. (For a general discussion on the transformation
of the units for mass, length and time we refer to Nariai and Ueno
\cite{7} and Dicke \cite{8}.) Eq. (\ref{355}) implies that the
over-all size of Einstein's gravitational constant 
and its dimension is determined by
the constant $K^{-1}$ with $\hph$, representing an elementary mass
ratio if the coupling constants $\gt ,~\gz$ or $\tilde\ga '$
in (\ref{348a}) were known, leading to a correction in the relation
between $\ka_E$ and $K$ as expressed by (\ref{355}).
The squared modulus of the scalar field --- being a constant
after Weyl-symmetry breaking --- enters the gravitational constant
$\ka_E$ in a manner reminiscent of the Brans-Dicke scalar-tensor
theory of gravitation \cite{3} although, as mentioned, the strength
of the gravitational coupling is determined in the presented broken
Weyl theory essentially by $K^{-1}$ having the dimension $[L/Energy]$.
For a more detailed investigation of this point see, however,
Subsection D below.
Furthermore, it is easy to show that taking the trace on both sides
of (\ref{354}) yields an identity after use of (\ref{311}) has
been made.

In the described situation where one considers the appearance of the
unit of length as originating from a $D(1)$- or Weyl-symmetry breaking
in a theory containing a universal scalar quantum field --- relating the
established length unit to the mass of this field --- the quantities
$\hbar$ and $c$ as well as the fine-structure constant $\a_F=\e^2/4\pi$
are regarded as universal constants being by definition unrelated
to the appearance of the length $\lp$. Summarizing we may say
that the scalar field in its $\Uem$ gauge invariant form
$\ph = \left({0\atop{\hph(x)}}\right)$, with $\hph(x)$ being a
constant, $\hph$, after $D(1)$-symmetry breaking, determines
according to Eqs. (\ref{347}) and (\ref{348a}) not only the $Z$- and
$W$-boson masses as well as the electron mass in a way described as
the ``Higgs phenomenon'' in the standard electroweak model --- which, in
fact, is just a choice of gauge called here the electromagnetic or nonlinear
gauge --- but affects also the gravitational coupling constant in
a Brans-Dicke-like manner with $\hat\Phi^2 = \hph^2$ playing the role 
of the real scalar field.\\

\noindent{\bf C. The Field Equations for the Scalar Field}

It is a surprising fact that the nonlinear $\p^4$-coupling term
proportional to  the constant $\b$ could be eliminated from the
energy-momentum tensor $\Theta_{\mn}^{(\p)}{}'$ due to Eq. (\ref{345})
following from the Weyl-symmetry breaking relation $D_\mu\Phi^2 = 0$.
In the electromagnetic gauge Eq. (\ref{345}) took the form of the
first equation in (\ref{346}) which then led, with $a=1$ and 
Eqs. (\ref{346b}) -- (\ref{348a}), to $\hat\Theta_{\mn}^{(\p)}{}'$
given in (\ref{351}).

In concluding this section let us now finally
study the field equation (\ref{34}) for $\p$ for the
case $a=1$, and $\ka_\mu =0$, $\hph = const$ according to (\ref{319}),
to see the influence of the $\p^4$-coupling on the dynamics of
the gauge and fermion fields in the nonlinear, i.e. electromagnetic,
gauge. To this end one has to 
compute $\hat{\appD}{}^\mu\hat{\appD}_\mu\ph$ using (B1) and (B3). 
It is then easy to show that (\ref{34}) is, in the
electromagnetic gauge, equivalent to the following $\Uem$ gauge
covariant equations:
\be
i{\gt\over{\sqrt{2}}}\,\hph\,\left[\bar\nabla^\mu\WW_\mu +
i\e\,\hat{\bar A}^\mu\WW_\mu - i\gz\,\ZZ^\mu\WW_\mu\right] +
2\tilde\ga '\,{mc\over\hbar}\,\hat{\bar e}_R \hat\nu_L = 0\/,
\label{359}
\ee
\be
i{\gt\over{\sqrt{2}}}\,\hph\,\left[\bar\nabla^\mu\WW^\da_\mu -
i\e\,\hat{\bar A}^\mu\WW^\da_\mu + i\gz\,\ZZ^\mu\WW^\da_\mu\right] -
2\tilde\ga '\,{mc\over\hbar}\,\hat{\bar\nu}_L \hat e_R = 0\/,
\label{360}
\ee
\be
-\left[ 2\tilde m_W^2\,\WW^\da{}^\mu\WW_\mu + 
\tilde m_Z^2\,\ZZ^\mu\ZZ_\mu\right] + {1\over 6}\bar R\hph^2 +
4\b\hph^4 +\left[{mc\over\hbar}\right]^2 \hph^2 +
{m_ec\over\hbar}\,\hat{\bar\psi}_e\hat\psi_e = 0\/,
\label{361}
\ee
\be
\bar\nabla^\mu\ZZ_\mu = 2\tilde\ga '\,{1\over{\gz\,\hph}}\,{mc\over\hbar}\,
i\left(\hat{\bar e}_L\hat e_R - \hat{\bar e}_R\hat e_L\right)\/.
\label{362}
\ee

Here (\ref{359}) corresponds to the upper components in (\ref{34})
when evaluated for $\ph =\left({0\atop\hph}\right)$,
Eq. (\ref{360}) is the adjoint equation of (\ref{359}),
while (\ref{361}) and (\ref{362}) are the real and imaginary parts
of the lower components in (\ref{34}), respectively. For comparison
with (\ref{346}) Eq. (\ref{361}) was multiplied by $\hph$. 
The constant $\b$ only enters Eq. (\ref{361}) 
which is seen to be identical to the first equation
of (\ref{346}) [prior to the use of (\ref{311}) and (\ref{348a})]
which was derived above.
Hence the only equation containing the term proportional to
$\b$ is, in fact, the algebraic equation Eq, (\ref{361}), and this
equation was used above to eliminate the $\b$-contribution
from the source terms in Einstein's equations. The elimination of
the $\b$-term is a particular consequence of the Weyl-symmetry breaking
by Eq. (\ref{32}) in this Weyl-electroweak theory.
Stated physically one may say: The energy represented by
the term proportional to $\b (\p^\da\p)^2$ in Eq. (\ref{349})
may be reexpressed by the mass terms appearing in Eq. (\ref{361}),
using also (\ref{311}), so that the constant
$\b$ disappears from the final equations. The field equation for $\p$
is thus, finally, turned into the set of linear differential equations
({\ref{359}), (\ref{360}) and (\ref{362}). \\

\noindent{\bf D. Determination of the Parameters of the Theory}

The following parameters appearing in the Lagrangean (\ref{33})
have already been fixed so far: $a = 1$ in (\ref{31}); $\tilde\a =0$
in $\LW$ [compare (\ref{228})], and $\dtt{}\!'=1/\e$ with
$\e = \sqrt{4\pi\a_F}=0.30282$ in (\ref{336}). Moreover, $\b$
disappeared from the dynamics after the Weyl-symmetry breaking as
was shown in the last subsection. The remaining parameters to be
determined are the following six quantities: The constants $\gt$
and $\gtt$ together with the Weinberg angle $\tw$ [compare (\ref{276})];
the constants $\hph$ and $K$; the Yukawa coupling constant $\tilde\ga '$
and, last not least, the universal length unit $\lp =\hbar/mc$ or
rather the reference mass $m = m_\p$. Besides the value for $\e$
already quoted (using $\a_F^{-1}=137.04$) we have the following
five experimental data at our disposal: $m_e=0.510999\, MeV/c^2,
m_Z=91.187\, GeV/c^2, m_W=80.41\, 
GeV/c^2, G_F=1.16639\cdot 10^{-5}\,GeV^{-2}$
and $\ka_E=2.076\cdot 10^{-48}g^{-1}cm^{-1}sec^2$.

Unfortunately it is not possible to decide uniquely what the actual length
scale $\lp$ is which is to be adopted as a universal unit in the theory.
We shall investigate two conceivable possibilities in somewhat greater detail:
(a) the mass of the $\p$-field is identical with the $Z$-boson mass,
i.e. $m=m^{(a)}=m_Z$, with
$\lp =\lp^{(a)}=\hbar/m_Zc=0.2164\cdot 10^{-15}\,cm$, and (b) the
mass of the $\p$-field is identical to the electron mass, i.e. $m=m^{(b)}=m_e$,
with $\lp =\lp^{(b)}=\hbar/m_ec=0.38610\cdot 10^{-10}\,cm$. As a
third possibility we only mention briefly the case when $\lp =\lp^{(c)}=1\,cm$
corresponding to $m=m^{(c)}=0.1973\cdot 10^{-4}\,eV/c^2$. This
possibility could be of interest in connection with a very small but
nonzero neutrino mass of order $m^{(c)}$ as the lower edge of the
fermion mass spectrum. Of course, this last choice, $\lp = 1\,cm$,
is completely ad hoc being included here only as an orientation.

For case (a) we find the following numerical values: $\cos\tw = 0.8818,~
|\gt |=0.65316$ [from $G_F$], $\gtt =0.34341$ and thus $|\gz |=0.73794$;
$\hph^2 = 7.345$ and $|\tilde\ga '|=0.2067\cdot 10^{-5}$ [from
$\hph = (m_e/m_Z)\tilde\ga '^{-1}$]. For $\hph >0$ the constants
$\gt$, $\gz$, and $\tilde\ga '$ must be positive, i.e. the absolute
signs in the quoted results are unnecessary.  Finally one has 
$K=0.7190/\ka_E$, i.e. $K$ is, for the case (a), essentially the 
inverse Einstein constant.

For the case (b) the constants $\gt, \gtt, \gz$ and $\tilde\ga '$
are the same as for the case (a) given above. However, now we have
$\hph^2 = 2.339\cdot 10^{11}$ due to $\hph = \tilde\ga '^{-1}$. This
leads, finally, to $K=2.565\cdot 10^{-11}/\ka_E$ being a factor of
the order of $(m_e/m_Z)^2$ smaller than in case (a). From this
it is apparent that the relative
contributions of $\hph^2$ and $K$ in the relation (\ref{355}) for
$\ka_E$ depends strongly on the unit of lengths adopted.

In concluding this subsection we remark that the question of the size
of the ``Higgs mass'' in the conventional formulation of the standard
model has turned in the present broken Weyl-electroweak theory into
the question of the relative contributions of $\hph^2$ and $K$ to
Einstein's gravitational constant and, correspondingly, into the
question of the actual size of the true elementary length scale to be
adopted in the theory. We, moreover, mention that in our determination of
the free parameters of the theory we used the observed electron, $Z_0$-
and $W^{\pm}$-boson masses disregarding radiative corrections.

The strength of the gravitational interaction is usually characterized
by the Planck length $l_{Planck}=(N\hbar /c^3)^\ha =1.616\cdot 10^{-33}\,cm$.
With this Einstein's gravitational constant $\ka_E$ may be written as
$\ka_E\cdot\hbar c=8\pi l_{Planck}^2=65.64\cdot 10^{-66}\,cm^2$.
However, writing finally Einstein's field equations (\ref{354}) relating
the contracted space-time curvature, i.e. the Einstein tensor $G_{\mn}$,
to the distribution of energy and momentum in a form independent of
a particular choice of a length unit yields
\be
G_{\mn} = \bar R_{\mn} - \ha g_{\mn} \bar R = {6\over{[\hph^2 + 1]}}\,T_{\mn}\,
\label{364}
\ee
where $G_{\mn}$ and $T_{\mn}$ are both to be measured in the same units of
an inverse length squared. The gravitational coupling in dimensionless form
is characterized in Eqs. (\ref{364}) by the constant $6/[\hph^2 + 1]$. For a
massless world, i.e. for $\hph =0$ in Eqs. (\ref{348a}), this dimensionless
coupling constant would at most be $6$; for the case (a) above it would be
$0.719$ --- i.e. of the order of unity as mentioned --- and for the
case (b) it would be $2.565\cdot 10^{-11}$.

\section{Discussion}

We investigated in this paper the semi-classical theory of a
scalar-isospinor field $\p$ coupled to the chiral fermion fields
$\pl$ and $\pr$ in the presence of the gauge fields $\ka_\mu$ (Weyl
vector fields) for the dilatation group $D(1)$, and the gauge fields
$A_\mu, Z_\mu, W_\mu$ and $W_\mu^\da$ for the electroweak gauge
group $\simG = SU(2)_W\times U(1)_Y$.
The dynamics of this originally massless and scaleless theory
was formulated in a Weyl space $W_4$ characterized by a family of metrics
$g_{\mn}$ and associated Weyl vector fields $\ka_\mu$, both determined
only up to Weyl transformations (2.3) and (2.4) corresponding
to conformal rescalings of the metric and the related transformations of
the Weyl vector fields, respectively. The gauge structure of the original
Weyl-electroweak theory (WEW theory) was given by the group
$G = SO(3,1)\otimes D(1)\otimes\simG$.

In order to investigate the appearance of nonzero masses and
establish a scale of lengths, $\lp$, in the theory which is associated
with the squared modulus, $\Phi^2 = \p^\da\p$, of Weyl weight $-1$
of the scalar field and, furthermore, derive field equations of
Einstein's type for the metric, we broke the Weyl-symmetry explicitly
by a term in the Lagrangean involving the scalar curvature $R$ of
the $W_4$ and a mass term for the scalar field. The idea here is
to establish an intrinsic length scale in an originally massless
and scaleless Weyl-symmetric theory by attributing this nonzero mass
and corresponding finite length unit to the scalar field $\p$. 
Then we studied how nonzero masses for
the various other interacting fields appear on the scene within the
framework of a broken gauge theory containing as a subsymmetry
the electroweak gauge symmetry which is known to contain many
features in accord with observation. After the Weyl-symmetry
breaking we finally obtain a $\Uem$ gauge covariant theory formulated
in a Riemannian space $V_4$. The reduction of the Weyl geometry
to a Riemannian geometry for the underlying space-time is
governed by a true symmetry breaking relation, $D_\mu\Phi^2 =0$,
implying that the $D(1)$ gauge field $\ka_\mu$ is ``pure gauge''
with the associated length curvature $f_{\mn}$ being zero and
the gauge symmetry with group $G$ reducing to a gauge symmetry
with the subgroup $G\,'= SO(3,1)\otimes\simG$. This is different
from the so-called spontanous symmetry breaking in the electroweak
sector of the theory which is better described as a choice of
gauge by singling out a particular point $\ph$ as origin in the
coset space $\simG\!/H$, with $\ph$ being invariant under
$H\equiv\Uem$, where $\simG\!/H$ is isomorphic to the scalar field
$\p$ (see Appendix A).

The transformation $\p\longrightarrow\ph$ is a gauge transformation
which reshuffles the fields by putting the theory in a form
possessing a residual $\Uem$ gauge freedom and exhibiting the
appearance of mass terms for the $\hat Z_\mu$-field, the $\hat W_\mu$-
and $\hat W_\mu^\da$-fields, and the electron field $\hat\psi_e$
without, however, reducing the connection and covariant derivatives
from a Lie\,$\simG$-valued form to a form characterized by a
corresponding expression associated with a subgroup of $\simG$.
This is contrary to the situation for the $D(1)$ or
Weyl-symmetry breaking described in this paper which, indeed, is a
true symmetry reduction $G\longrightarrow G\,'$ in the sense of the
theorem quoted at the end of Sec. II ~leading to the appearance
of the length scale $\lp$ in the theory freezing at the same
time the squared modulus $\hat\Phi^2$ of the scalar field to
the constant $\hph^2$. Now the question arises: What is the true
nature of this scalar ``field'' $\hph$ which enters the $Z_0$- and
$W^{\pm}$-boson masses relating them and the electron mass to
the established length scale and, furthermore, enters the gravitational
constant in Einstein's field equations for the metric in the
$V_4$ limit in a manner comparable to a Brans-Dicke field
$\hph^2$. We try to answer this question in the following way: The
$\p$-field, as it appears in the broken Weyl theory, is a
{\it vehicle for symmetry breaking}. $\p$ is not a matter field of
the usual type which would also possess a particle interpretation
in a fully quantized theory. For this reason it is very unlikely
that this field, being in the $V_4$ limit reduced to a constant
responsible for the mass generation of the gauge boson and
charged fermion fields, would actually show up as a particle
in high energy processes. It is a field necessary to establish
a scale in a theory. Now the next question arises: What is the
actual size of this scale? Of course, here we have to rely on
observation. Since the measured masses $m_Z$ and $m_W$ are
of the order of $100\, GeV/c^2$ it is suggestive to assume
that the mass scale established by the $\p$-field after
Weyl-symmetry breaking is of this same order. The corresponding
length $\lp$ would thus be of the order of
$\lp \approx 0.2\cdot 10^{-15}\,cm$ [case (a) in Subsection III D].
However, this identification, although reasonable, is
not compelling. Regardless of whether one fixes the mass $m$ at
the scale of $100\, GeV/c^2$ --- case (a) above --- or identifies
this $\p$-mass with the much smaller electron mass --- case
(b) above --- the huge difference of the observed
masses for the $Z_0$- and $W^{\pm}$-bosons, on the one hand, and the
electron mass, on the other hand, is mainly due to the Yukawa
coupling constant $\tilde\ga '$ [compare Eqs. (\ref{348a})].
A point of particular interest in the presented broken Weyl theory,
however, is that the special choice of the unit of lengths also
affects the dimensionless coupling constant appearing in the field
equations (\ref{364}) for gravity.

\vspace{1cm}
\noindent{\bf ACKNOWLEDGEMENT}

\noindent I thank Heinrich Saller for numerous discussions.

\newpage
\begin{appendix}
\section
{Coset Representation of $\phi$}

The coset representation of $\p$ is related to the proper disentanglement
of the various $U(1)$ phase groups involved. The field $\p$ transforms
under $G=SO(3,1)\otimes D(1)\otimes \simG$ corresponding
to a representation with spin zero, Weyl weight $w(\p)=-\ha$, isospin $I=\ha$,
and hypercharge $Y=\ha$. The $D(1)$ factor of $G$ affects the modulus
$\Phi = \sqrt{\p^\da\p}$, while the electroweak gauge group
\be
\simG\;= SU(2)_W \times U(1)_Y
\label{A1}
\ee
determines the orientation of the isospin degrees of freedom
and the $U(1)_Y$ phase. We denote by $U(\bar g)$, with $\bar g\in\simG$,
the $2\times 2$ representation of $\simG$ operating on $\p$. We
parametrize the elements of $\simG$ by $\bar g = \bar g(b_a,\b)$
with $b_a;~a=1,2,3$, yielding a parametrization of $SU(2)_W$, and with
$\b$ parametrizing the hypercharge transformations. (The angle $\b$ 
should not be confused with the constant $\b$ used in Eq. (\ref{228})
and in the main text.) Thus the $U(1)_Y$
transformations are given by
\be
U\Bigl(\bar g(0,\b)\Bigr) = \left(\begin{array}{cc} e^{i\gtt Y\b}&0 \\
                       0&e^{i\gtt Y\b} \\ \end{array}\right)
\label{A2}
\ee
with $Y=\ha$. Mathematically speaking, the transformations (\ref{A2})
are transformations of $U(2)$ which may be decomposed into the
direct product
\be
U(\bar g(0,\b) = U(1)_+ \otimes U(1)_-~,
\label{A3}
\ee
where the groups $U(1)_{\pm}$ are generated by $\ha(1\pm\tau_3)$,
respectively, i.e. by the electromagnetic charge $\q=\ha(1+\tau_3)$
[compare Eq. (\ref{265})] and by
\be
\q_o = \ha (1 - \tau_3)\/.
\label{A4}
\ee
The decomposition of the original weak hypercharge and isospin transformations
into $\q$ and $\q_0$ contributions (of which, as shown below, only the
$\q$ contributions survive) corresponds to physically measurable
situations yielding the coupling to the $A_\r$-fields (electromagnetic
effects) and the coupling to the $Z_\r$-fields (weak neutral effects).

We now like to introduce a representation of $\p$ which is characterized
in terms of the cosets $\simG\!/H$, where $H$ is the {\it electromagnetic}
subgroup of $\simG$, i.e. $H=\Uem =U(1)_+\subset\simG$. We thus write
\be
\p = U(\gp )~\ph  \qquad {\rm with} \qquad \q~\ph =0\/.
\label{A5}
\ee
Here $U(\gp)$ is an element of $\simG\!/H$ parametrized by $\p$, and
\be
\ph = \left({0\atop \hat\vphi_0}\right)\/,
\label{A6}
\ee
with $\hat\vphi_0$ being a real field, denotes the origin of the coset space.
$\ph$ is invariant under the electromagnetic gauge group $\Uem =U(1)_+$
[the stability group $H$] with transformations $h\in H$ given by
\be
U(h(\a)) = e^{-i{e\over\hbar c}\q\a} = \left(\begin{array}{cc}
           e^{-i{e\over\hbar c}\a} & 0\\ 0 & 1 \\ \end{array}\right)\/,
\label{A7}
\ee
where the minus sign in the exponential is adopted for conventional reasons.
Due to the splitting (\ref{A3}) of the hypercharge transformations
and the invariance of $\ph$ by the contributions $U(1)_+$ generated
by the charge $\q$, the transformation $U(\gp)$ --- which could be
called a ``boost'' generating $\p$ from the fixed state $\ph$ --- is
seen to be given by the following element of $SU(2)_W\otimes U(1)_-$ :
\be
U(\gp ) = U\Bigl(\bar g(b_a(\p)\Bigr)~e^{{i\over 2}\gtt\q_o\b(\p)}\/,
\label{A8}
\ee
with $\q_o$ as given by (\ref{A4}). Here the first factor on the r.-h. side
is an element of $SU(2)_W$ parametrized by $b_a(\p)$, and the second
factor is the transformation
\be
\left(\begin{array}{cc} 1 & 0 \\ 0 & e^{{i\over 2}\gtt\b(\p)} \\
\end{array}\right) \in U(1)_-
\label{A9}
\ee
with hypercharge phase angle $\b(\p)$. It is easy to show that one
can express the r.-h. side of (\ref{A8}) in terms of the components
of $\p$ in the following manner [compare (\ref{222})]:
\be
U(\gp ) = {1\over {\hat\vphi_0}}\left(\begin{array}{cc}
         ~\vphi^*_0~e^{{i\over 2}\gtt\b(\p)} & ~\vphi_+ \\
         -\vphi^*_+~e^{{i\over 2}\gtt\b(\p)} & ~\vphi_0 \\ \end{array}
\right)\/,
\label{A10}
\ee
with $\hat\vphi_0 =\sqrt{\p^\da\p} = \sqrt{|\vphi_+|^2 + |\vphi_0|^2}$
expressing the invariance of the modulus $\Phi$ of $\p$ under
``boosts'' parametrized in terms of $\simG\!/H$.
Identifying $\p$ with the coset $U(\p)\cdot H\equiv U(\gp)\cdot H$
we may now take the simpler matrix $U(\p)$ defined by
\be
U(\p) = {1\over {\hat\vphi_0}}\left(\begin{array}{cc}
       ~\vphi^*_0 & ~\vphi_+ \\
       -\vphi^*_+ & ~\vphi_0 \\ \end{array}
\right)\/,\quad {\rm with} \quad \det U(\p)=1
\label{A11}
\ee
as a coset representative instead of $U(\gp)$ and write Eq. (\ref{A5}) as
\be
\p = U(\p)~\ph\/.
\label{A12}
\ee

The $\simG$-transformation of $\p$ or, more exactly, the gauge
transformation $U(\bar g(x))$ representing a change of section on the
bundle $E$ [see (\ref{224})] given by (the argument $x$ is suppressed)
\be
\p\;' = U(\bar g)~\p \/,
\label{A13}
\ee
together with the coset representation (\ref{A12}) of $\p$ as well as of
$\p '$ associated with the stability group $H=\Uem$ now yields the
following decomposition of an arbitrary transformation
$U(\bar g)$ into boosts parametrized by $\p =\p (x)$ and $\p '=\p '(x)$,
respectively, and a stability group transformation
characterized by an angle $\a$ depending on $\p (x)$ and
on $\bar g=\bar g(x)$ which is written for short as $\a (\p ',\p)$ :
\be
U(\bar g) = U(\p ')~e^{-i{e\over \hbar c}\q \a (\p ',\p)}~
            U^{-1}(\p)\/.
\label{A14}
\ee
Here the subgroup transformation
\be
U(h(\p ',\p)) = e^{-i{e\over \hbar c}\q \a (\p ',\p)} =
              \left(\begin{array}{cc}
              e^{-i{e\over{\hbar c}}\a(\p ',\p)} & 0 \\ 0 & 1 \\
              \end{array}\right)
\label{A15}
\ee
could be called the ``Wigner rotation'' for the electroweak theory
or the ``little group transformation'' at the origin $\ph$ in the coset space
$\simG\!\!/H$ which is associated with the transformation $U(\bar g)=
U(\bar g(b_a,\b))$ sending $\p$ into $\p '$. The transformation
(\ref{A15}) with angle $\a(\p ',\p)$ is in similar contexts 
usually called the {\it nonlinear realization} of a gauge transformation
of the group $\simG$ on the stability subgroup $H$ of $G$ \cite{5}.
In the standard model, however, this terminology is not
used and one speaks instead of a symmetry breaking by the vacuum
expectation value of the scalar field $\p$ having the form (\ref{A6}).
In the present case, with $\simG$ possessing
the product structure (\ref{A1}), one finds for the angle $\a (\p ',\p)$
by direct computation for the transformations $\bar g=\bar g(b_a,\b)$
the results
\begin{eqnarray}
&&\qquad {\rm for} \qquad \bar g(0,0) ~~:~~~~~ \a(\p ,\p) = 0\/, \nonumber\\
&&\qquad {\rm for} \qquad \bar g(0,\b)~~:~{e\over {\hbar c}}\a(\p ',\p) =
                                        -\gtt\b\/,\nonumber \\
&&\qquad {\rm for} \qquad \bar g(b_a,0)~~:~{e\over {\hbar c}}\a(\p ',\p) = 0\/.
\label{A16}
\end{eqnarray}
The last line in (\ref{A16}) implies that there are no residual $SU(2)_W$
gauge transformations left on the stability
subgroup $H=\Uem$. After transforming to the origin $\hat\p$
in $\simG\!/H$ only one gauge degree of freedom remains which
is of the form (\ref{A15}) with $\a(\p ',\p)$ given by (\ref{A16})
together with the corresponding transformation rule for the
electromagnetic potentials $A_\mu$ [see Appendix B].
The result for the hypercharge transformation
$\bar g(0,\b)$ quoted in (\ref{A16}) follows also directly from the
determinant of the transformation (\ref{A14}).

\section
{The Electromagnetic Gauge}

We call the gauge obtained by realizing the transformations
$U(\bar g(b_a,\b)$ of $\simG$ in terms of transformations $U(h(\p ',\p))$
of the electromagnetic subgroup $H=U(1)_+$ of $\simG$ the
electromagnetic or nonlinear gauge [compare Eqs. (\ref{A14}) and (\ref{A15})].
To characterize this gauge, which we shall denote by a hat,
the scalar field $\p$ of the theory is used: As described in Appendix A,
$\p$ takes the form $\ph =\left({0\atop\hph}\right)$ in this gauge, and
the residual gauge transformations are the transformations of the
stability subgroup $U(1)_+=\Uem$ leaving $\ph$ invariant.

It is of particular interest to determine the form the gauge potentials
$A^a_\mu$ and $B_\mu$ take in the electromagnetic gauge, i.e. determine
$\hat A^a_\mu$ and $\hat B_\mu$ and their residual gauge freedom given
by the transformations $h(\p ',\p)$. Let us to this end rewrite the
$G$-covariant derivative (\ref{227}) of $\p$ in terms of the spherical
components $A^-_\mu =W_\mu ,~A^+_\mu =W^\da_\mu$ and $A^3_\mu$
[compare (\ref{274})] and express $A^3_\mu$ and $B_\mu$ in terms of
$A_\mu$ and $Z_\mu$ with the help of (\ref{270}). One finds with
$\tau_\pm = {1\over {2\sqrt 2}}(\tau_1\pm i\tau_2)$:
\be
\appD_\mu \p = \left[D_\mu + i\gt \left(W_\mu\tau_+ + W^\da_\mu\tau_-\right)
             +i\,\e\,{e\over{\hbar c}}\q A_\mu + 
              i\gz\left(\ha\tau_3 - \sin^2\tw\cdot
             \q\right)Z_\mu\right] \p \/.
\label{B1}
\ee

Performing now a gauge transformation with $U^{-1}(\p)$, mapping
$\p$ into $\ph$, the covariant derivative (\ref{B1}) of $\p$ is mapped into
\be
\hat{\appD}_\mu \ph = \left[D_\mu + i\gt\hat W_\mu \tau_+ - {i\over 2}
                      \gz\hat Z_\mu\right] \ph \/,
\label{B2}
\ee
where we have used the fact that $\q\ph =0$, $\tau_-\ph =0$ and
$\ha\tau_3\ph = -\ha\ph$. We rewrite this for later use in terms of
isospinor components, i.e.
\be
\hat{\appD}_\mu\ph = \left({0\atop{D_\mu\hph}}\right) + i \left(
{{{1\over{\sqrt 2}}\,\gt\hph\,\hat W_\mu}\atop{-{1\over 2}\,\gz\hph\,\hat Z_\mu}}
\right)\/,
\label{B3}
\ee
where $D_\mu\hph =\pa_\mu\hph +\ha\ka_\mu\hph$ is the Weyl-covariant
derivative of the real field $\hph$. From the residual transformation
with $U(h(\p ',\p))$ given by (\ref{A15}) one now concludes that
\be
\left(\hat{\appD}_\mu\ph\right)' = \left(\begin{array}{cc}
                   e^{-i{e\over{\hbar c}}\a(\p ',\p)} & 0 \\ 0 & 1 \\
                   \end{array}\right)~\hat{\appD}_\mu\ph
\label{B4}
\ee
is the nonlinear subgroup transformation rule for the $G$-covariant
derivative of $\ph$. One sees at once from the split form
(\ref{B3}) that (\ref{B4}) implies for the residual (electromagnetic)
$U(1)_+$ gauge transformations of the potentials $\hat Z_\mu$,
$\hat W_\mu$ and $\hat W_\mu^\da$ the behaviour
\begin{eqnarray}
\hat Z_\mu ' &=& \hat Z_\mu \/, \label{B5} \\
\hat W_\mu ' &=& e^{-i{e\over{\hbar c}}\a(\p ',\p)}~\hat W_\mu\/, \label{B6} \\
\hat W_\mu^\da{}' &=& e^{+i{e\over{\hbar c}}\a(\p ',\p)}~\hat W_\mu^\da \/.
        \label{B7}
\end{eqnarray}
For the electromagnetic potentials $\hat A_\mu$ applies the usual 
rule given by [see Eq. (\ref{B13a}) below]
\be
\hat A_\mu ' = \hat A_\mu + \pa_\mu\a(\p ',\p)\/.
\label{B8}
\ee
It is clear from the definitions that $\hat A_\mu$ and $\hat Z_\mu$
are real vector fields.

In concluding this appendix we observe with respect to the theorem
quoted at the end of Sect. II that the connection in the present case
does {\it not} reduce from a $Lie \simG$-valued to a
$Lie H$-valued form. In fact, we have for the $Lie \simG$-valued
gauge potentials with Eqs. (\ref{270}), (\ref{274}) and (\ref{276}):
\begin{eqnarray}
\Gt_\mu &=& \gt \ha\tau_a A^a_\mu + \gtt \ha {\bf 1} B_\mu \nonumber \\
        &=& \left(\begin{array}{cc}
      \e\,\bar A_\mu + \gz\left[\ha - \sin^2\tw\right]Z_\mu &
      \gt{1\over{\sqrt 2}}W_\mu \\ \gt{1\over{\sqrt 2}}W^\da_\mu &
      -\ha\gz Z_\mu \\ \end{array}\right)
\label{B9}
\end{eqnarray}
with $\bar A_\mu = \eh A_\mu$.
In this notation the covariant derivative is written
$\appD_\mu\!\p = D_\mu\p +i\Gt_\mu\!\p$. The transformation to the
electromagnetic gauge with $U^{-1}(\p)$ yields
\be
\hat{\Gt}_\mu = U^{-1}(\p)\Gt_\mu U(\p) - iU^{-1}(\p)~\pa_\mu U(\p)
\label{B10}
\ee
which is still $Lie \simG$-valued and hence, according to the theorem of
Sect. II, the $\simG$ gauge symmetry does not reduce but is nonlinearly
realized.

From (\ref{B10}) we compute for the fields $\hat W_\mu$ and $\hat Z_\mu$
appearing in (\ref{B2}) the expressions:
\begin{eqnarray}
{\gt\over{\sqrt 2}}\hat W_\mu &= \Biggl[\left(\e\,\bar A_\mu +
  \gz [\ha - \sin^2\tw ]Z_\mu\right)&\vv_0\vv_+ + {\gt\over{\sqrt 2}}W_\mu (\vv_0)^2
       -{\gt\over{\sqrt 2}}W^\da_\mu (\vv_+)^2 \nonumber \\
  &\qquad\qquad + \ha\gz Z_\mu\vv_0\vv_+\Biggr] (\hph)^{-2}&
       -i\left[\vv_0\pa_\mu\!\left({\vv_+\over\hph}\right)
                   -\vv_+\pa_\mu\!\left(\vv_0\over\hph\right)\right](\hph)^{-1} \/,
\label{B11} \\
-\ha\gz\hat Z_\mu &= \Biggl[\left(\e\,\bar A_\mu +
  \gz [\ha -\sin^2\tw ]Z_\mu\right)&|\vv_+|^2
                                      + {\gt\over{\sqrt 2}}W_\mu \vv^*_+\vv_0
         +{\gt\over{\sqrt 2}}W^\da_\mu\vv^*_0\vv_+ \nonumber \\
  &\qquad\qquad - \ha\gz Z_\mu |\vv_0|^2\Biggr] (\hph)^{-2}&
      -i\left[\vv^*_+\pa_\mu\!\left({\vv_+\over\hph}\right)
                  +\vv^*_0\pa_\mu\!\left({\vv_0\over\hph}\right)\right](\hph)^{-1}\/.
\label{B12}
\end{eqnarray}
For $\hat{\bar A}_\mu =\eh\hat A_\mu $, by inserting also 
$\hat Z_\mu$ from (\ref{B12}), one finds the relation:
\begin{eqnarray}
\e\,\hat{\bar A}_\mu +& &\gz[\ha - \sin^2\tw]\hat Z_\mu =
  \Biggl[\left(\e\,\bar A_\mu + \gz [\ha - \sin^2\tw ]Z_\mu\right)|\vv_0|^2
   -{\gt\over{\sqrt 2}}W_\mu\vv_0\vv^*_+ \nonumber \\
& & -{\gt\over{\sqrt 2}}W^\da_\mu\vv^*_0\vv_+
                       - \ha\gz Z_\mu |\vv_+|^2\Biggr] (\hph)^{-2}
    -i\left[\vv_0\pa_\mu\!\left(\vv^*_0\over\hph\right)
                    +\vv_+\pa_\mu\!\left(\vv^*_+\over\hph\right)\right](\hph)^{-1}\/.
\label{B13}
\end{eqnarray}

After the transformation (\ref{B10}) has been carried out the residual
gauge freedom is given by the transformations $U(h(\p ',\p))$ of the
electromagnetic subgroup $U(1)_+ = \Uem$, i.e. by
\be
\hat{\Gt}_\mu{}'= U(h(\p ',\p))\,\hat{\Gt}_\mu \,U^{-1}(h(\p ',\p)) -
                  i\,U(h(\p ',\p))\,\pa_\mu U^{-1}(h(\p ',\p))\/,
\label{B13a}
\ee
with $U(h(\p ',\p))$ as defined in (\ref{A15}) with phase angle (\ref{A16}). For
the potentials $\hat A_\mu ', \hat Z_\mu ', \hat W_\mu '$ and
$\hat W_\mu '^\da$ this yields at once the relations (\ref{B5}) -- (\ref{B8}).

We finally quote the form of the fermion fields after transformation to
the origin in $\simG\!/H$:
\be
\hat\pl = U^\da(\p)\,\pl \qquad {\mbox {\rm and}} \qquad \hat\pr = \pr \/,
\label{B14}
\ee
implying that
\be
\hat\nu_L = (\hph)^{-1}[\vv_0\,\nu_L-\vv_+\,e_L]\/,\quad
       \hat e_L = (\hph)^{-1}[\vv^*_+\,\nu_L+\vv^*_0\,e_L]\/,\quad \hat e_R = e_R\/.
\label{B15}
\ee
The residual $\Uem$ gauge transformations of these fields are
\be
\hat\nu_L {}' = e^{-i{e\over{\hbar c}}\a(\p ',\p)}~\hat\nu_L\/,\quad
     \hat e_L {}' = \hat e_L\/,\quad \hat e_R {}' = \hat e_R \/,
\label{B16}
\ee
and correspondingly for the adjoint fields $\hat{\bar\nu}_L, \hat{\bar e}_L$
and $\hat{\bar e}_R$. The Yukawa coupling (\ref{226}) of $\p$ and the
fermion fields may in this notation, together with $U^\da(\p)\,\p = \ph$,
be written as
\be
\tilde\ga\left\{(\bar\pl\p)\pr + \bar\pr(\p^\da\pl)\right\}\equiv
\tilde\ga\left\{(\hat{\bar\pl}\ph)\hat\pr + \hat{\bar\pr}(\ph^\da\hat\pl)\right\}
= \tilde\ga\,\hph ~\left\{\hat{\bar e}_L \hat e_R + \hat{\bar e}_R \hat e_L\right\}\/.
\label{B17}
\ee
The r.-h. side of (\ref{B17}) may be written as 
$\tilde\ga\,\hph\,\hat{\bar\psi}_e\hat\psi_e$, with $\hat\psi_e$ denoting
the electron field transformed to the origin in $\simG\!/H$, showing
that the Yukawa coupling represents, in effect, an electron mass term.

\end{appendix}


\end{document}